\documentclass[12pt]{article}
\usepackage{epsfig}
\usepackage{amssymb}
\usepackage{amsmath}
\usepackage{amsfonts}
\usepackage{graphicx}
\usepackage{mathrsfs}
\DeclareMathAlphabet{\mathscrbf}{OMS}{mdugm}{b}{n}
\usepackage{mathabx}
\usepackage[dvips]{color}
\usepackage{multirow}
\usepackage{calc}
\usepackage{accents}

\makeatletter
\newcommand{\shorteq}{\mathrel{\mkern0.2mu\mathpalette\shorteq@\relax\mkern0.2mu}}
\newcommand{\shorteq@}[2]{\scalebox{0.5}[1]{$\m@th#1=$}}

\newcommand{\longeq}[1]{\mathrel{\mathpalette\longeq@{#1}}}
\newcommand{\longeq@}[2]{%
  \begingroup
  \sbox\z@{$\m@th#1=$}%
  \ifdim#2<\wd\z@
    \resizebox{#2}{\height}{\box\z@}%
  \else
    \ifdim#2<3\wd\z@
      \hbox to #2{$\m@th#1=\hss=\hss=\hss=$}%
    \else
      \hbox to #2{$\m@th#1=\cleaders\hbox to 0.2\wd\z@{\hss$#1=$\hss}\hfil=$}%
    \fi
  \fi
  \endgroup
}
\makeatother

\newcommand{\bnabla}{\boldsymbol{\nabla}}

\newcommand{\R}{\mathbb{R}}
\newcommand{\C}{\mathbb{C}}

\newcommand{\fm}{\mathfrak{m}}
\newcommand{\fn}{{\mathfrak{n}}}

\newcommand{\fz}{\mathfrak{z}}

\newcommand{\fK}{\mathfrak{K}}

\newcommand{\bfe}{\mathbf{e}}

\newcommand{\bk}{\mathbf{k}}

\newcommand{\bfr}{\mathbf{r}}

\newcommand{\bE}{\mathbf{E}}

\newcommand{\bH}{\mathbf{H}}
\newcommand{\bI}{\mathbf{I}}

\newcommand{\bM}{\mathbf{M}}

\newcommand{\bS}{\mathbf{S}}

\newcommand{\cH}{\mathcal{H}}

\newcommand{\cM}{\mathcal{M}}

\newcommand{\cO}{\mathcal{O}}
\newcommand{\cP}{\mathcal{P}}
\newcommand{\cQ}{\mathcal{Q}}
\newcommand{\cR}{\mathcal{R}}

\newcommand{\cT}{\mathcal{T}}
\newcommand{\cU}{\mathcal{U}}

\newcommand{\cX}{\mathcal{X}}

\newcommand{\be}{\begin{equation}}
\newcommand{\ee}{\end{equation}}
\newcommand{\bea}{\begin{eqnarray}}
\newcommand{\eea}{\end{eqnarray}}
\newcommand{\nn}{\nonumber}

\newcommand{\ed}{\end{document}}

\newcommand{\bi}{\begin{itemize}}
\newcommand{\ei}{\end{itemize}}

\newcommand{\bce}{\begin{center}}
\newcommand{\ece}{\end{center}}

\newcommand{\sS}{\mathscr{S}}
\newcommand{\sT}{\mathscr{T}}

\newcommand{\bPsi}{{\boldsymbol{\Psi}}}

\newcommand{\bcM}{{\boldsymbol{\cM}}}

\newcommand{\bcH}{{\boldsymbol{\cH}}}
\newcommand{\bcU}{{\boldsymbol{\cU}}}

\newcommand{\for}{{\mbox{\rm for}}}

\begin{document}

\title{Scattering of TE and TM waves and quantum dynamics generated by non-Hermitian Hamiltonians}

\author{Farhang Loran\thanks{E-mail address: loran@iut.ac.ir}
~and Ali~Mostafazadeh\thanks{Corresponding author, E-mail address:
amostafazadeh@ku.edu.tr}\\[6pt]
$^*$Department of Physics, Isfahan University of Technology, \\ Isfahan 84156-83111, Iran\\[6pt]
$^\dagger$Departments of Mathematics and Physics, Ko\c{c} University,\\  34450 Sar{\i}yer,
Istanbul, T\"urkiye}

\date{ }
\maketitle

\begin{abstract}
The study of the scattering of electromagnetic waves by a linear isotropic medium with planar symmetry can be reduced to that of their TE and TM modes. For situations where the medium consists of parallel homogeneous slabs, one may use the standard transfer matrix technique to address the scattering problem for these modes. We extend the utility of this technique to inhomogeneous permittivity and permeability profiles by proposing a dynamical formulation of the scattering of TE and TM waves in which the transfer matrix for the medium is given in terms of the evolution operator for an effective non-unitary quantum system. This leads to a system of dynamical equations for the reflection and transmission amplitudes. Decoupling these equations we reduce the solution of the scattering problem for TE and TM modes to that of an initial-value problem for a Riccati equation. We discuss the application of this observation in identifying media that do not reflect TE or TM waves with given wavenumber and incidence angle. 
\end{abstract}

\section{Introduction}

Maxwell's equations describing the propagation of transverse electric (TE) waves by an isotropic nonmagnetic linear medium with planar symmetry may be reduced to the Helmholtz equation,
	\be
	\partial_x^2\psi(x)+k^2[\fn(x)^2-\sin^2\theta]\psi(x)=0,
	\label{HH-eq-TE}
	\ee
where $\fn(x)$ is the refractive index of the medium, and $k$ and $\theta$ are respectively the wavenumber and incidence angle. Because this equation has the same structure as the time-independent Schr\"odinger equation,
	\be
	[-\partial_x^2+v(x)]\psi(x)=k^2\psi(x),
	\label{sch-eq}
	\ee
we can employ tools of quantum scattering theory in one dimension to deal with the scattering of TE waves by inhomogeneities of such a medium. Principal examples are the scattering and transfer matrices \cite{muga-review,sanchez,tjp-2020}.  

Suppose that $v(x)$ is a short-range potential, i.e., it decays to zero faster that $1/x$ as  $x\to\pm\infty$. Then the solutions of (\ref{sch-eq}) have the following asymptotic behavior.
	\be
	\psi(x)\to A_\pm e^{ik x}+B_\pm e^{-ik x}~~~~\for~~~~x\to\pm\infty,
	\label{asymptot-psi}
	\ee	
where $A_\pm$ and $B_\pm$ are $x$-independent complex coefficients that determine the amplitudes of the right-going and left-going waves (with respect to the standard orientation on the $x$ axis), respectively. The scattering and transfer matrices of $v(x)$ are respectively the $2\times 2$ complex matrices, $\bS$ and $\bM$, that satisfy
	\begin{align}
	&\bS\left[\begin{array}{c}
	A_-\\
	B_+\end{array}\right]=\left[\begin{array}{c}
	A_+\\
	B_-\end{array}\right],
	\label{S=}\\
	&\bM\left[\begin{array}{c}
	A_-\\
	B_-\end{array}\right]=\left[\begin{array}{c}
	A_+\\
	B_+\end{array}\right],
	\label{M=}
	\end{align}
and are independent of $A_\pm$ and $B_\pm$, \cite{muga-review,sanchez,tjp-2020}.

In scattering setups, the source of the incident wave resides at either of $x=-\infty$ or $x=+\infty$. These correspond to solutions of (\ref{sch-eq}) with $A_-\neq 0=B_-$ or $B_+\neq 0=A_-$. We call them left-incident and right-incident waves and denote them by $\psi^l$ and $\psi^r$, respectively. Using the superscripts $l$ and $r$ to label the corresponding amplitudes, $A_\pm$ and $B_\pm$, we can define the left and right reflection $R^{l/r}$ and transmission $T^{l/r}$ amplitudes of $v(x)$ by
	\begin{align}
	&R^l:=\frac{B^l_-}{A^l_-},
	&&T^l:=\frac{A^l_+}{A^l_-},
	&&R^r:=\frac{A^r_+}{B^r_+},
	&&T^r:=\frac{B^r_-}{B^r_+}.
	\label{RT-def}
	\end{align}
Combining these with (\ref{S=}) and (\ref{M=}) and making use of the fact that the Wronskian of any pair of solutions of \eqref{sch-eq} is constant, we find \cite{tjp-2020},
	\begin{align}
	&\det\bM=1, 
	&&T^l=T^r=\frac{1}{M_{22}}=S_{11}=S_{22},
	\label{T=}\\
	&R^l=-\frac{M_{21}}{M_{22}}=S_{21},
	&&R^r=\frac{M_{12}}{M_{22}}=S_{12},
	\label{R=}
	\end{align}
where $M_{ij}$ and $S_{ij}$ are the entries of $\bM$ and $\bS$, respectively \cite{tjp-2020,Springer-book-2018}. Because $T^l=T^r$, we use $T$ to refer to the left- and right-transmission amplitudes.  Eqs.~(\ref{T=}) and \eqref{R=} imply
	\begin{align}
	&\bM=\frac{1}{T}\left[\begin{array}{cc}
	T^2-R^lR^r & R^r\\
	-R^l & 1\end{array}\right],
	&&\bS=\left[\begin{array}{cc}
	T & R^r\\
	R^l & T\end{array}\right].
	\label{M=S=}
	\end{align}

Transfer matrix has an important practical advantage over the scattering matrix {known as its (de)composition property. To describe it, consider expressing $v$ as the sum of $n$ short-range potentials $v_1,v_2,\cdots,v_n$ give by 
	\begin{align}
	&v_1(x):=\left\{\begin{array}{cc}
	v(x)&\for~x\leq a_1,\\
	0 &{\rm otherwise},\end{array}\right.
	&&v_n(x):=\left\{\begin{array}{cc}
	v(x)&\for~x> a_{n-1},\\
	0 &{\rm otherwise},\end{array}\right.
	\nn\\
	&v_j(x):=\left\{\begin{array}{cc}
	v(x)&\for~a_{j-1}<x\leq a_j,\\
	0 &{\rm otherwise},\end{array}\right.
	&&j\in\{2,3,4,\cdots,n-1\},
	\nn
	\end{align} 
where $a_1,a_2,\cdots,a_{n-1}$ are arbitrary real numbers such that $a_1<a_2<\cdots<a_{n-1}$. Let $\bM_1,\bM_2,\cdots,\bM_n$ denote  the transfer matrices of the potentials $v_1$, $v_2,\cdots,v_n$, respectively. Then, the transfer matrix $\bM$ of $v$ satisfies \cite{sanchez,tjp-2020,Springer-book-2018},
	\be
	\bM=\bM_n\bM_{n-1}\bM_{n-2}\cdots\bM_1.
	\label{compose}
	\ee}%
It is important to notice that this equation holds for any choices of $n$ and $a_1,a_2,\cdots,a_n$. 

For a finite-range piecewise continuous potential, we can choose $a_1$ and $a_n$ such that $v(x)=0$ for $x<a_1$ and $x>a_n$, take $n$ sufficiently large, and make sure that the discontinuities of $v$ coincide with some of the $a_2,a_3,\cdots,a_{n-1}$. Then, $\bM_1=\bM_2=\bI$, and we can
approximate $v_2,v_3,\cdots,v_{n-1}$ respectively with barrier potentials of hight $v(a_2), v(a_3),\cdots,v(a_{n-1})$ whose transfer matrices $\bM_2,\bM_3,\cdots\bM_{n-1}$ admit closed-form analytic formulas \cite{ap-2014}. Substituting these in \eqref{compose} we find an approximate expression for $\bM$ which we can improve by taking larger values of $n$.

The (de)composition property \eqref{compose} which is the key ingredient of the above scheme of slicing the potential into pieces and computing its transfer matrix in terms of those of its slices is the main reason for the introduction of the transfer matrix \cite{jones-1941,abeles,thompson}, its numerous applications \cite{teitler-1970,yeh,abrahams-1980,ardos-1982,levesque,sheng-1996,griffiths,prl-2009},  and generalizations \cite{berreman-1972,pendry-1984,pendry-1990a,pendry-1990b,pendry-1996,pereyray-1998,wang-2001,katsidis-2002,Shukla-2005,Hao-2008,li-2009,zhan-2013,ap-2016,pra-2021,pra-2023}.

Ref.~\cite{ap-2014} reveals an intriguing connection between the transfer matrix $\bM$ and the dynamics of a certain non-unitary two-level quantum system. Specifically, it constructs a $2\times 2$ non-Hermitian matrix Hamiltonian $\bcH(\tau)$ whose evolution operator $\bcU(\tau,\tau_0)$ satisfies $\bM=\bcU(-\infty,\infty)$. 	 Here the evolution parameter $\tau$ is a constant multiple of $x$. In particular, we can identify it with $x$ in which case $\bcH(x)$ takes the form \cite{tjp-2020}, 
	\be
	\bcH(x):=\frac{v(x)}{2k}\left[\begin{array}{cc}
	1 & e^{-2ikx}\\
	-e^{2ikx} & -1\end{array}\right].	
	\label{H=1D}
	\ee
	
Let us recall that with $x$ playing the role of ``time'', the evolution operator $\bcU(x,x_0)$ satisfies
	\begin{align}
	&i\partial_x\,\bcU(x,x_0)=\bcH(x)\bcU(x,x_0),
	&&\bcU(x,x_0)=\bI,
	\label{sch-eq-U}
	\end{align}	
where $x_0$ represents the initial ``time'', and $\bI$ is the $2\times 2$ identity matrix. Because $\bcH(x)$ is a non-stationary Hamiltonian, we do not have an explicit expression for $\bcU(x,x_0)$. We can however expand it in a Dyson series and identify it with the time-ordered exponential $\bcH(x)$;
	\begin{align}
	\bcU(x,x_0)&=\bI+\sum_{n=1}^\infty(-i)^n
	\int_{x_0}^x\!\!dx_n \int_{x_0}^{x_n}\!\!dx_{n-1}\cdots\int_{x_0}^{x_2}\!\!dx_1\:
	\bcH(x_n)\bcH(x_{n-1})\cdots\bcH(x_1)
	\nn\\
	&=\sT\exp[-i\int_{x_0}^xdx'\:\bcH(x')],
	\nn
	\end{align}
where $\sT$ stands for the time-ordering operator \cite{weinberg}. In particular, we have 
	\be
	\bM=\sT\exp\left[-i\int_{-\infty}^\infty dx\:\bcH(x)\right].
	\label{M=exp}
	\ee
Note also that because $\bcH(x)$ is non-Hermitian, it generates a non-unitary evolution.\footnote{This is consistent with the fact that, in general,  $\bM$ is not a unitary matrix.}
	
Since time-ordered exponential of traceless matrix Hamiltonians have unit determinant and $\bcH(x)$ is clearly traceless, Eq.~(\ref{M=exp}) provides a simple proof of the identity $\det\bM=1$. Furthermore we can use this equation and the well-known composition property of evolution operators in quantum mechanics to give a simple proof of the (de)composition property (\ref{compose}), \cite{tjp-2020}. 

An immediate consequence of the above connection between the transfer matrix and non-unitary quantum dynamics is the derivation of dynamical equations for the reflection and transmission amplitudes, $R^{l/r}$ and $T$, \cite{ap-2014}. These have provided the impetus for developing an inverse-scattering scheme for devising optical systems with desired scattering properties at a single pre-assigned frequency \cite{ap-2014,pra-2014a,pra-2014b}. The subsequent work on the subject has revealed an interesting relationship between the semi-classical scattering and adiabatic approximation \cite{jpa-2014a,jpa-2014b},  paved the way towards the development of a transfer matrix for long-range scattering potentials \cite{jpa-2020b}, led to an effective method of computing the coefficients of the low-frequency series expansions of the reflection and transmission amplitudes \cite{jmp-2021,jpa-2021}, and provided a road map for devising a fundamental concept of transfer matrix for potential scattering in two and three dimensions \cite{pra-2021}. 

The purpose of the present article is to extend the dynamical formulation of potential scattering developed in Ref.~\cite{ap-2014} to the scattering of TE and TM waves by inhomogeneities of a general (possibly magnetic) isotropic linear medium with planar symmetry.  In Sec.~\ref{S2} we define the transfer matrix for TE and TM waves and discuss its basic properties. In Sec.~\ref{S3} we extend the domain of  validity of Eq.~\eqref{M=exp} to TE and TM waves by deriving an analog of the matrix Hamiltonian (\ref{H=1D}) for these waves. In Sec.~\ref{S4}, we obtain dynamical equations for the corresponding reflection and transmission amplitudes and show that it reduces to a single Riccati equation. In Sec.~\ref{S5} we explore the application of this equation for identifying optical systems that do not reflect TE or TM waves with given wavenumber and incidence angle. In Sec.~\ref{S6} we present a summary of our findings and concluding remarks.

\section{Transfer matrix for TE and TM waves}
\label{S2}

Consider a charge-free linear and isotropic scattering medium $\sS$ with planar symmetry. Choosing a Cartesian coordinate system in which $\sS$ has translational symmetry along the $y$ and $z$ axes, we can express the permittivity $\varepsilon$ and permeability $\mu$ of $\sS$ as functions of $x$.\footnote{In general they also dependent on the wavenumber of the incident wave through a dispersion relation. This does not however affect the analysis of this paper, for they hold for a single value of $k$.} Since we wish to study the scattering of electromagnetic waves due to the inhomogeneities of $\sS$, we suppose that as $x\to\pm\infty$, $\varepsilon(x)$ and $\mu(x)$ tend to the permittivity and permeability of vacuum, $\varepsilon_0$ and $\mu_0$, faster than $1/x$.{\footnote{{The results of this article apply to situations that $\sS$ is immersed in a homogenous background medium filling the space in place of vacuum. In this case $\varepsilon_0$ and $\mu_0$ should be replaced by the permittivity and permeability of the background, respectively.}}} In terms of the relative permittivity and permittivity of $\sS$, i.e.,
	\begin{align}
	&\hat\mu(x):=\frac{\mu(x)}{\mu_0}, &&
	\hat\varepsilon(x):=\frac{\varepsilon(x)}{\varepsilon_0},\nn
	\end{align}
we can state this condition as follows.
	\be
	\lim_{x\to\pm\infty}x[\hat\varepsilon(x)-1]=\lim_{x\to\pm\infty}x[\hat\mu(x)-1]=0.
	\label{asymptot}
	\ee
Furthermore, we demand that there is a positive number $C$ such that the first and second derivatives of $\varepsilon(x)$ and $\mu(x)$ exist for $|x|\geq C$ and decay to zero faster than $1/x$ as $x\to\pm\infty$, i.e.,	
	\be
	\lim_{x\to\pm\infty}x\,\partial_x^j\hat\varepsilon(x)=\lim_{x\to\pm\infty}x\,\partial_x^j\hat\mu(x)=0~~~\for~~~j\in\{1,2\}.
	\label{asymptot-d}
	\ee
		
\begin{figure}
        \begin{center}
        \includegraphics[scale=.2]{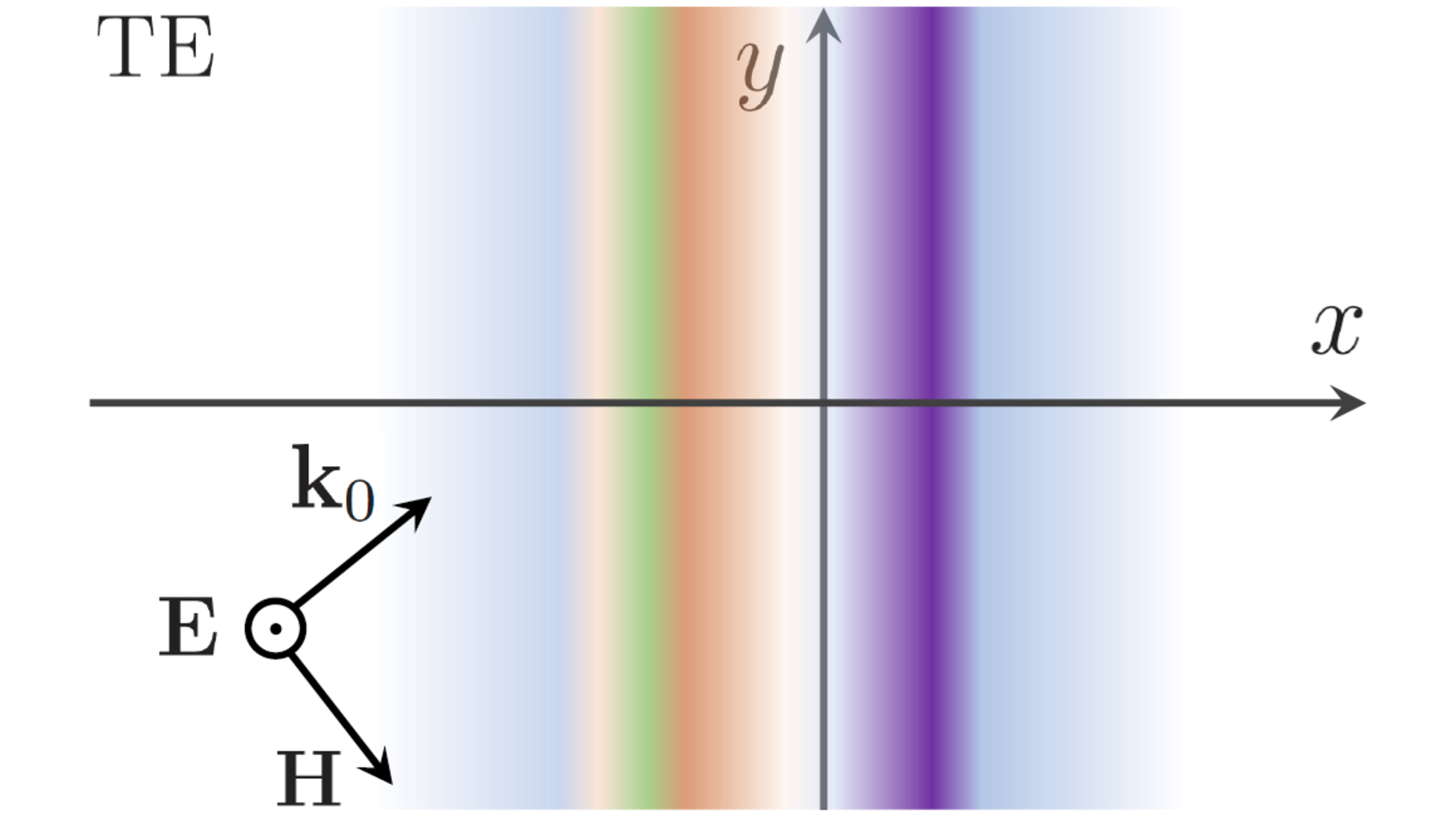}~~~~~~~~~
        \includegraphics[scale=.2]{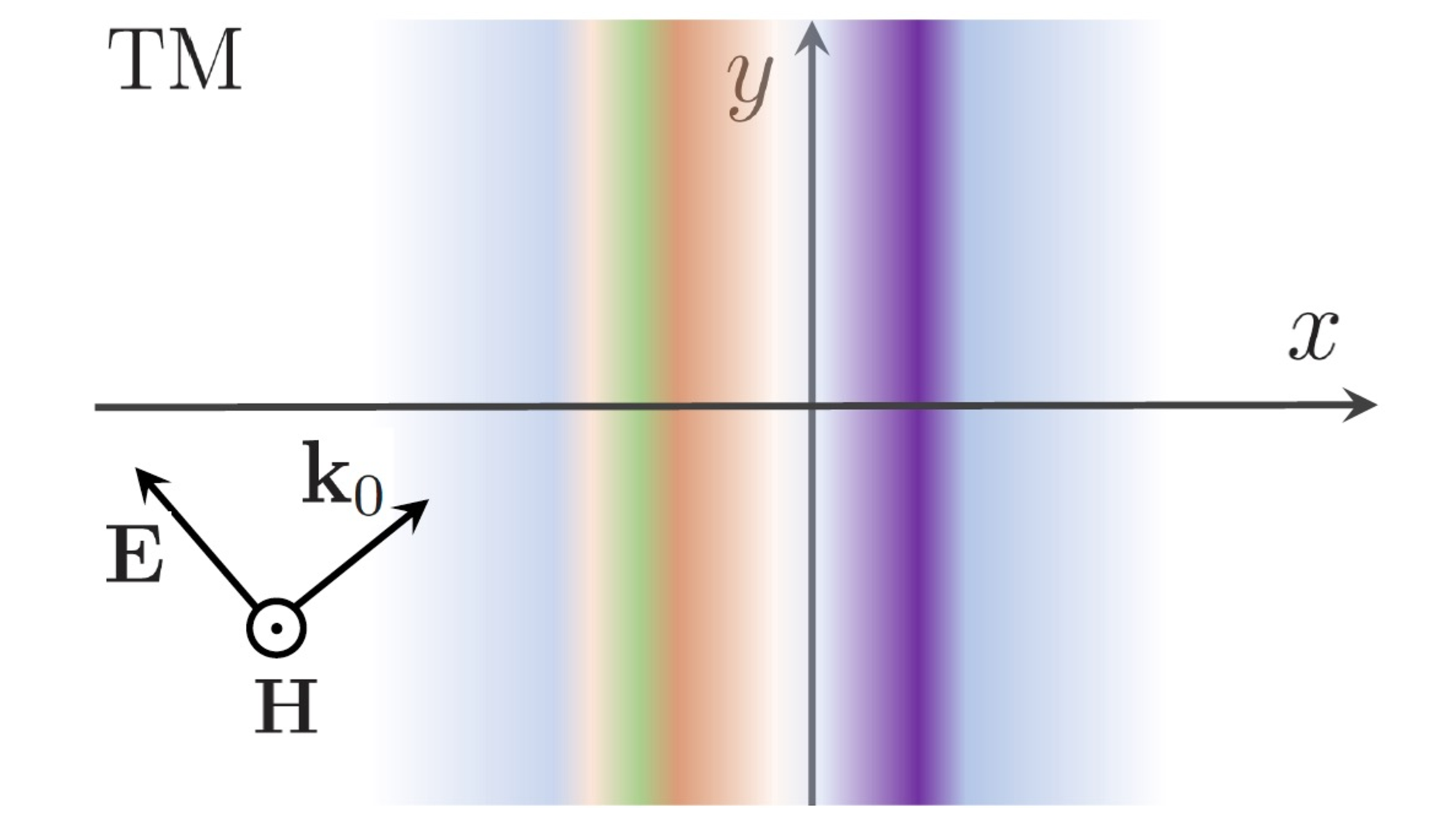} 
        \caption{{Schematic views of TE and TM waves propagating in an isotropic medium with planar symmetry along the $y$ and $z$ directions. The latter is denoted by $\odot$.}}
        \label{fig1}
        \end{center}
        \end{figure}

Given a time-harmonic TE (respectively TM) wave propagating in $\sS$, we can align the $y$ and $z$ axes of our coordinate system such that the electric field of the TE wave (magnetic field of the TM wave) lies along the $z$ axis while the incident wave vector $\bk_0$ is parallel to the $x$-$y$ plane. See Fig.~\ref{fig1}. 
Using $\theta$ to denote the angle between $\bk_0$ and the positive $x$ axis (incidence angle), we have
	\begin{align}
	&\bk_0=k_x\hat\bfe_x+k_y\hat\bfe_y,
	&&k_x:=k\cos\theta,
	&&k_y:=k\sin\theta,
	\label{kkk}
	\end{align}
where $\hat\bfe_u$ is the unit vector along the positive $u$ axis, $u\in\{x,y,z\}$, and $k$ is the incident wavenumber. 

Next, we recall that Maxwell's equations for time-harmonic electromagnetic fields have the form,
	\begin{align}
    	&\bnabla\cdot(\varepsilon\,\bE)=0, && \bnabla\cdot(\mu\,\bH)=0,
    	\nn\\
	&\bnabla\times\bE=i\omega\mu\,\bH,
	&&\bnabla\times\bH=-i\omega\varepsilon\,\bE,     	
	\nn
    	\end{align}
where $e^{-i\omega t}\bE(\bfr)$ and $e^{-i\omega t}\bH(\bfr)$ are respectively the electric and magnetic fields, and $\omega$ is the angular frequency. TE and TM waves correspond to the following solutions of these equations.
    \begin{align}
    &{\rm TE:}~\left\{\begin{aligned}
    &\bE=e^{ik\sin\theta\, y}\,\psi(x) \hat{\bfe}_z, \\
    &\bH=[c\,\mu(x)]^{-1}e^{ik\sin\theta\,  y}
    \left[\sin\theta\,\psi(x)\,\hat{\bfe}_x+ik^{-1}\partial_x\psi(x)\,\hat{\bfe}_y\right],
    \end{aligned}\right.
    \label{TE=}
    \\[6pt]
    &{\rm TM:}~\left\{\begin{aligned}
    &\bE=-[c\,\varepsilon(x)]^{-1}e^{ik\sin\theta\,  y}
    \left[\sin\theta\,\psi(x)\hat{\bfe}_x+ik^{-1}\partial_x\psi(x)\hat{\bfe}_y\right],\\
    &\bH=e^{ik\sin\theta\, y}\psi(x)\hat{\bfe}_z,
    \end{aligned}\right.
    \label{TM=}
    \end{align}
where $c:=(\varepsilon_0\mu_0)^{-1/2}=\omega/k$ is the speed of light in vacuum,  $\psi$ is a bounded solution of 
	\begin{align}
	&\alpha(x)\,\partial_x\!\left[\alpha(x)^{-1}\partial_x\psi(x)\right]+
	k^2\left[\fn(x)^2-\sin^2\theta\right]\psi(x)=0,
	\label{HH-eq}
	\end{align}
$\fn$ is the (complex) refractive undex of $\sS$ which satisfies $\fn^2=\hat\varepsilon\hat\mu$, and 
	\be
	\alpha:=\left\{\begin{array}{ccc}
	\hat\mu&\for&\mbox{TE waves},\\
	\hat\varepsilon&\for&\mbox{TM waves}.
	\end{array}\right.
	\label{alpha=}
	\ee

Using the second equation in \eqref{kkk}, we can express \eqref{HH-eq} as
	\begin{align}
	&\alpha(x)\,\partial_x\!\left[\alpha(x)^{-1}\partial_x\psi(x)\right]+\fK^2\,\tilde\fn(x)^2\psi(x)=0,
	\label{HH-eq-2}
	\end{align}
where 
	\begin{align}
	&\fK:=|k_x|=k|\cos\theta|,
	&\tilde\fn(x):=\pm|\sec\theta|\sqrt{\fn(x)^2-\sin^2\theta},
	\label{tn=}
	\end{align}
and the $\pm$ in the expression for $\tilde\fn$ is to be chosen so that the real parts of $\fn$ and $\tilde\fn$ have the same sign \cite{ap-2016}. This is positive for ordinary matter and negative for negative-index metamaterial \cite{veselago,pendry-2000,smith-2008,shalaev}. {Equation~\eqref{HH-eq-2} is known as the Bergmann's equation in acoustics where it describes the propagation of time-harmonic pressure waves in a compressible fluid \cite{Bergmann,Martin}.}
	
\begin{figure}
        \begin{center}
        \includegraphics[scale=.17]{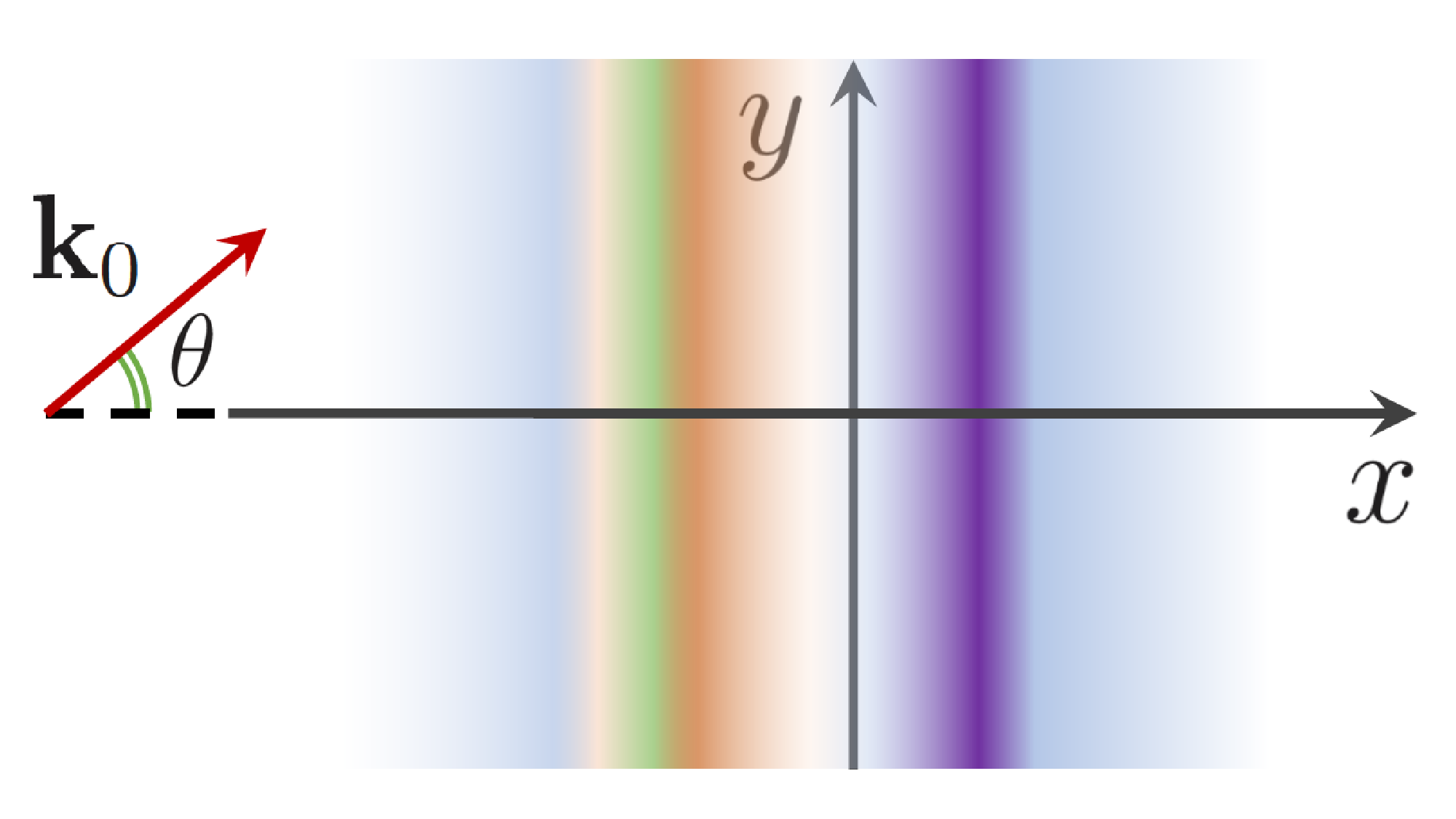}~~~~~~~~~
        \includegraphics[scale=.17]{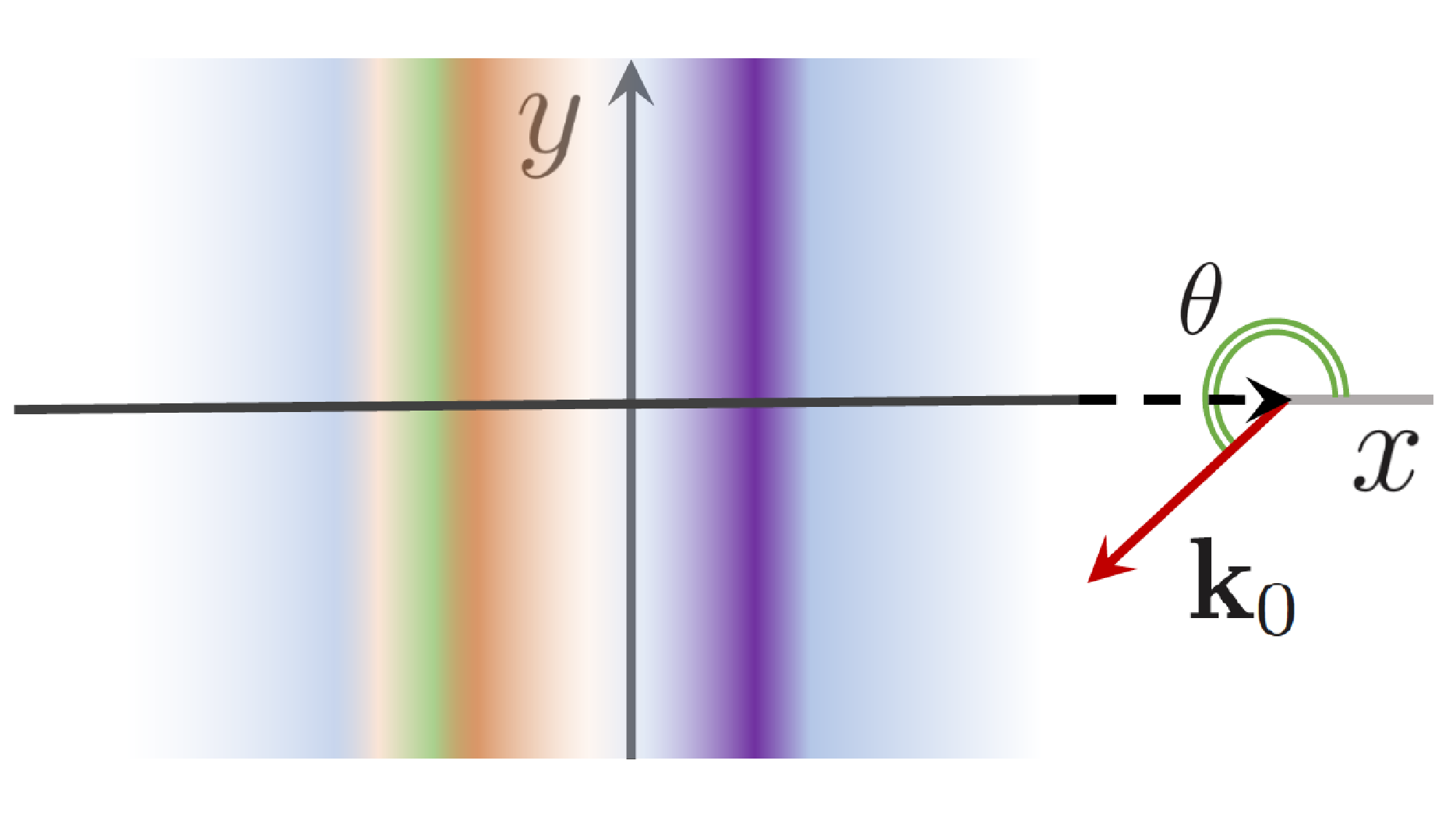} 
        \caption{{Wave vectors $\bk_0$ shown as a red arrow for a left-incident wave (on the left) and a right-incident wave (on the right). The incidence angles $\theta$ for  
left-incident and right-incident waves respectively satisfy $-90^\circ<\theta<90^\circ$ and $90^\circ<\theta<270^\circ$.}}
        \label{fig2}
        \end{center}
        \end{figure}

In view of (\ref{asymptot}) and (\ref{tn=}), for $x\to\pm\infty$, $\hat\varepsilon(x)^{\pm 1}-1$, $\hat\mu(x)^{\pm 1}-1$, and consequently $\tilde\fn(x)^2-1$ decay to zero faster than $1/x$. We can use this observation together with \eqref{asymptot-d} to infer that the solutions $\psi$ of (\ref{HH-eq-2}) fulfill \eqref{asymptot-psi} with $\fK$ replacing $k$, i.e., given such a solution there are complex coefficients $A_\pm$ and $B_\pm$ such that
	\be
	\psi(x)\to A_\pm e^{i\fK x}+B_\pm e^{-i\fK x}~~~~\for~~~~x\to\pm\infty.
	\label{asymptot-psi-K}
	\ee	
This relation enables us to identify the transfer matrix $\bM$ of the medium $\sS$ for the TE and TM waves with the the $2\times 2$ matrix $\bM$ that satisfies \eqref{M=} and is independent of the coefficients $A_\pm$ and $B_\pm$. Furthermore, it allows us to employ the same definitions for the left- and right-incident waves and the left and right reflection and transmission amplitudes for $\sS$, namely (\ref{RT-def}). {In particular, we use the terms left-incident and right-incident waves for incidet waves whose sources are respectively located at $x=-\infty$ and $x=+\infty$. This means that, as show in Fig.~\ref{fig2}, the incidence angle for a left-incident (respectively right-incident) wave satisfies $-90^\circ<\theta<90^\circ$ (respectively $90^\circ<\theta<270^\circ$).}

It turns out that the argument given in \cite{tjp-2020} to establish the identity, $\det\bM=1$, applies also for the transfer matrix of $\sS$ for the TE and TM waves, and we can express $R^{l/r}$ and $T^{l/r}$ in terms of the entries of $\bM$ using (\ref{T=}) and \eqref{R=}. We can also verify that $\bM$ possesses the (de)composition property \eqref{compose}.

The similarity between the transfer matrix of quantum scattering in one dimension and the transfer matrix of $\sS$ for the TE and TM waves has its limitations. This stems from the fact that unlike time-independent Schr\"odinger equation (\ref{sch-eq}), Eq.~(\ref{HH-eq-2}) involves a term proportional to $\partial_x\psi$. If $\alpha$ is constant, which is the case for TE waves propagating in a nonmagnetic medium, this term vanishes, (\ref{HH-eq-2}) reduces to \eqref{HH-eq-TE}, and we recover the full analogy with quantum scattering defined by the Schr\"odinger equation~\eqref{sch-eq} with $k$ to be replaced by $\fK$ and the potential $v$ given by
	\be
	v(x):=\fK^2[1-\tilde\fn(x)^2]=k^2[1-\fn(x)^2].
	\label{optical-v}
	\ee
This is not the case for TE (respectively TM) waves scattered by the inhomogeneities of an isotropic medium with variable permeability (respectively permittivity).

If $\alpha$ is a piecewise constant function, the term proportional to $\partial_x\psi$ in (\ref{HH-eq-2}) disappears in regions where $\alpha(x)$ is constant. To determine the solution of (\ref{HH-eq-2}) on the whole real line, however, we must impose Maxwell's boundary conditions at the discontinuities of $\alpha$. {For the system we consider, these demand the $y$ and $z$ components of $\bE$ and $\bH$ to be continuous at these points \cite{Jackson}. In view of \eqref{TE=} and \eqref{TM=}, this means that $\psi(x)$ and $\alpha(x)^{-1}\partial_x\psi(x)$ must be} continuous functions of $x$ at these points and consequently in $\R$. This is in contrast with the solutions of the Schr\"odinger equation (\ref{sch-eq}) which are required to be continuous and have a continuous derivative in $\R$.

\begin{figure}
        \begin{center}
        \includegraphics[scale=.22]{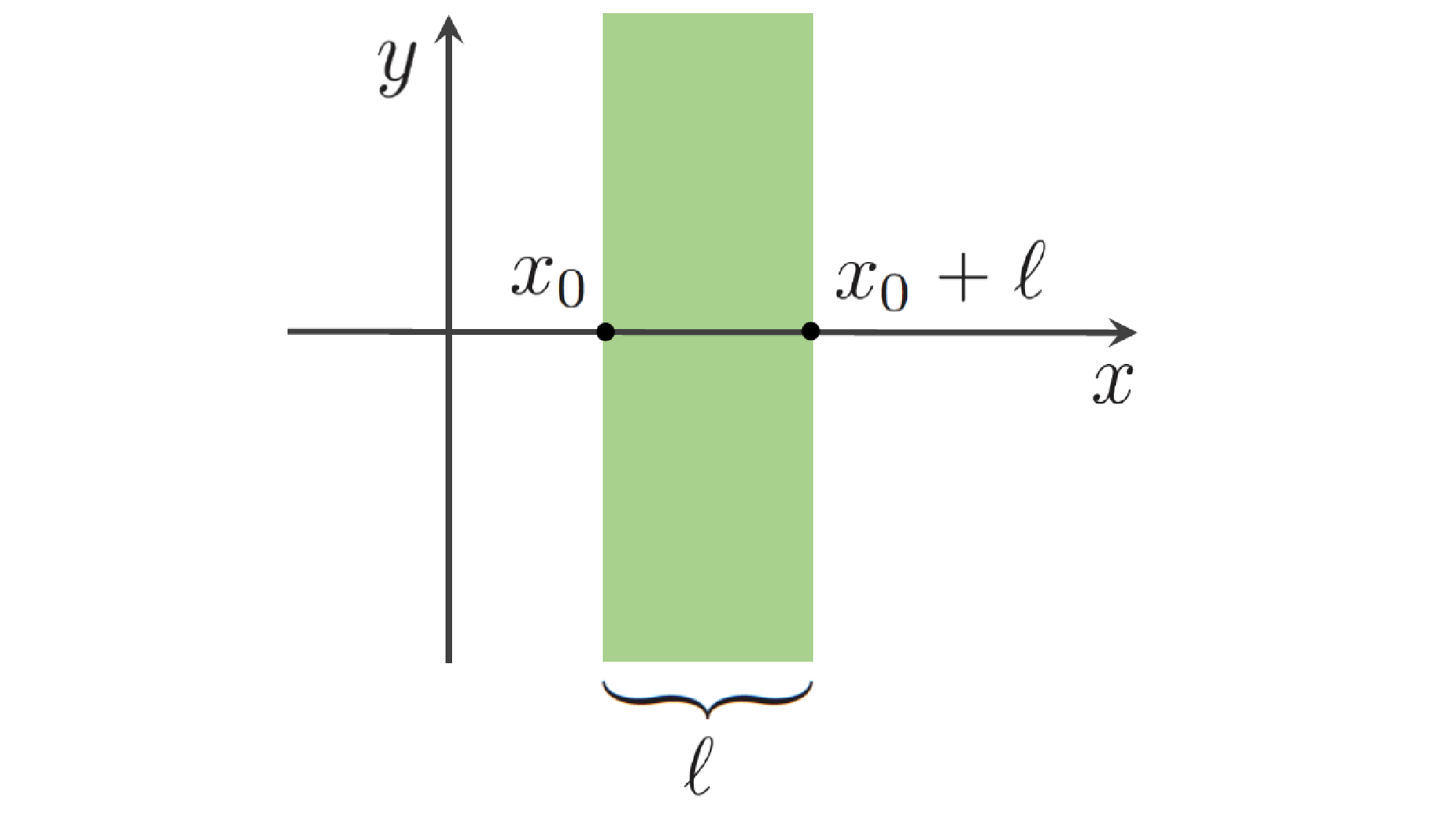}~~
        \includegraphics[scale=.22]{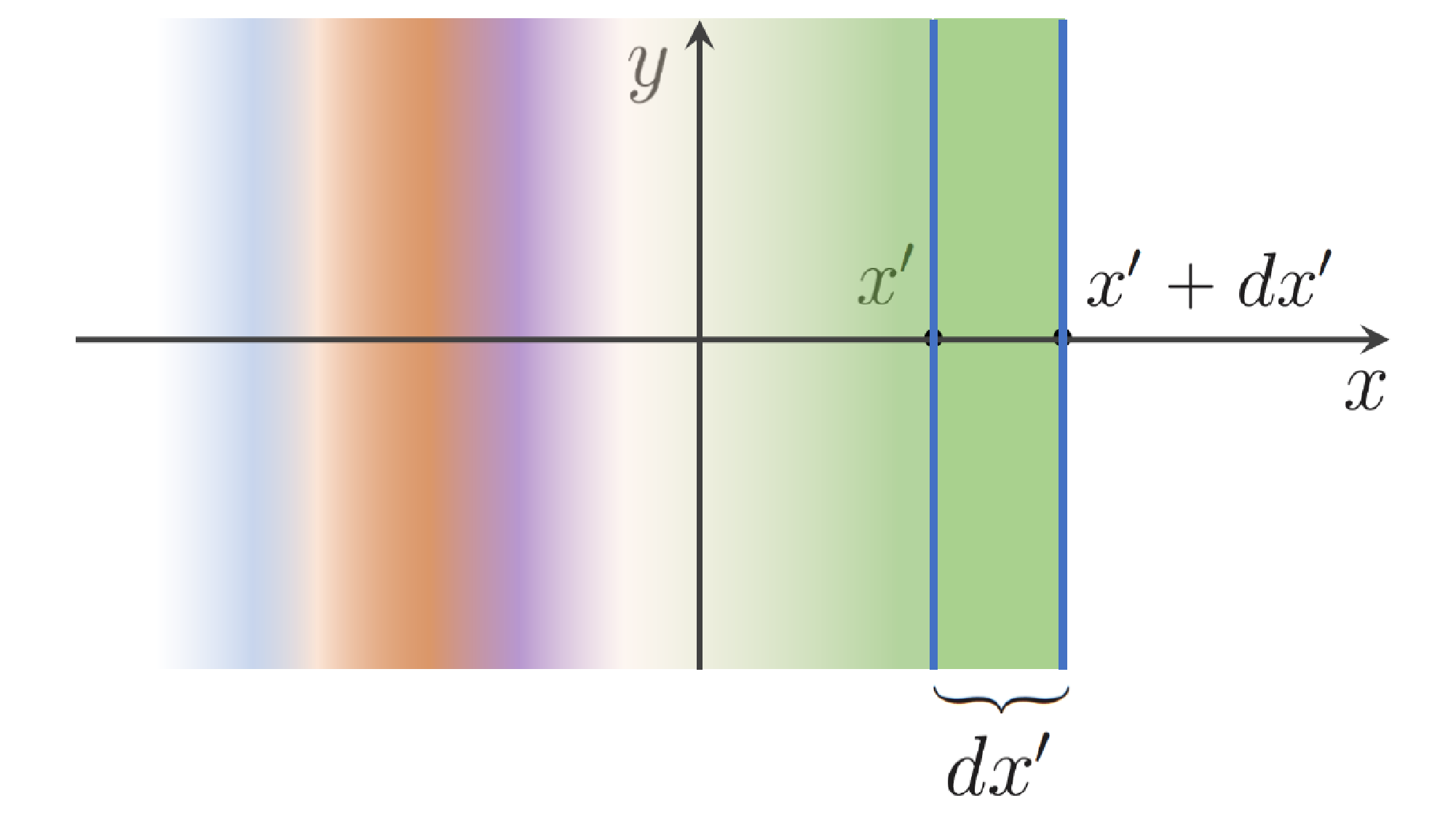} 
        \caption{{Schematic views of a homogeneous planar slab lying between the planes $x=x_0$ and $x=x_0+\ell$ on the left, and the truncated inhomogeneous medium $\sS_{x'+dx'}$ occupying the half-space given by $x\leq x'+dx'$ on the right. The latter consists of a slab of infinitesimal thickness $dx'$ attached to the truncated medium $\sS_{x'}$ which fills the half-space $x\leq x'$.}}
        \label{fig3}
        \end{center}
        \end{figure}

As a simple example, consider cases where the scattering medium is a homogenous planar slab made of an isotropic linear material (or metamaterial) placed in vacuum. See the left-hand panel in Fig.~\ref{fig3}.
Ref.~\cite{ap-2016} studies the scattering of TE and TM waves for such a slab and uses the equivalence of the matching conditions at its boundaries with the presence of certain point interactions to determine slab's transfer matrix. For a slab of thickness $\ell$ that occupies the region given by ${x_0}\leq x\leq {x_0}+\ell$, this calculation gives
	\be
	\bM_{\rm slab}=\left[\begin{array}{cc}
	(\cos\fm+i\fn_+\sin\fm)e^{-i\fK\,\ell} & i\fn_-\sin\fm\, e^{-i\fK(2{x_0}+\ell)}\\[6pt]
	-i\fn_-\sin\fm\, e^{i\fK(2{x_0}+\ell)} & (\cos\fm-i\fn_+\sin\fm)e^{i\fK\,\ell}\end{array}\right],
	\label{M-slab}
	\ee
where ${x_0}$ is a real parameter, and
	\begin{align}
	&\fm:=\fK\,\ell\,\tilde\fn,
	&&\fn_\pm:=
	\frac{1}{2}\left(\frac{\tilde\fn}{\alpha}\pm\frac{\alpha}{\tilde\fn}\right).
	\label{m-npm}
	\end{align}

\section{Quantum dynamics for TE and TM wave scattering}
\label{S3}

Let $\sS$ be the general isotropic medium with planar symmetry that we consider in Sec.~\ref{S2}, 
and for each ${x'}\in\R$, use $\sS_{x'}$ to label the isotropic medium whose relative permittivity and relative permeability are respectively given by
	\begin{align}
	&\hat\varepsilon_{x'}(x):=\left\{
	\begin{array}{ccc}
	\hat\varepsilon(x)&\for& x\leq {x'},\\
	1 &\for&x>{x'},\end{array}\right.
	&&\hat\mu_{x'}(x):=\left\{
	\begin{array}{ccc}
	\hat\mu(x)&\for& x\leq {x'},\\
	1 &\for&x>{x'}.\end{array}\right.
	\label{truncated-1}
	\end{align}
These equations imply
	\begin{align}
	&\lim_{{x'}\to\infty}\hat\varepsilon_{x'}(x)=\hat\varepsilon(x),
	&&\lim_{{x'}\to\infty}\hat\mu_{x'}(x)=\hat\mu(x),
	&&\sS=\lim_{{x'}\to\infty}\sS_{x'},
	\label{limit}
	\end{align}
where by the last equation we mean that we recover $\sS$ from $\sS_{x'}$ by letting ${x'}$ tend to $\infty$. We refer to $\sS_{x'}$ as the medium obtained by truncating $\sS$ at $x'$. See Fig.~\ref{fig3}.
	
Next, consider a slab of infinitesimal thickness $d{x'}$ that is bounded by the planes given by $x={x'}$ and $x={x'}+d{x'}$, and having the following relative permittivity and permeability profiles.
	\begin{align}
	&\hat\varepsilon_{\rm slab}(x):=\left\{
	\begin{array}{cc}
	\hat\varepsilon({x'})&\for~{x'}\leq x\leq {x'}+d{x'},\\
	1 &{\rm otherwise},\end{array}\right.
	&&\hat\mu_{\rm slab}(x):=\left\{
	\begin{array}{cc}
	\hat\mu({x'})&\for~{x'}\leq x\leq{x'}+d{x'},\\
	1 &{\rm otherwise}.\end{array}\right.
	\nn
	\end{align}
According to (\ref{M-slab}) and (\ref{m-npm}), the slab's transfer matrix has the form,
	\be
	\bM_{\rm slab}=\bI-i\bcH({x'})d{x'},
	\label{M-slab=2}
	\ee
where we have ignored quadratic and higher order terms in powers of $dx'$ and introduced,
	\begin{align}
	&\bcH(x):=\fK\left[\begin{array}{cc}
	-\fm_+(x)+1 
	&-\fm_-(x)\,e^{-2i\fK\, x}\\[6pt]
	\fm_-(x)\,e^{2i\fK\, x}&\fm_+(x)-1
	\end{array}\right].
	\label{Ha=}\\[6pt]
	&\fm_\pm(x):=\frac{\tilde\fn(x)^2\pm\alpha(x)^2}{2\alpha(x)}= 
	\frac{\sec^2\theta[\fn(x)^2-1]\pm\alpha(x)^2+1}{2\alpha(x)}.
	\label{mmp=}
	\end{align}
	
If we denote the transfer matrix of $\sS_{x'}$ by $\bcM({x'})$, we can use the composition property~(\ref{compose}) and Eq.~\eqref{M-slab=2} to establish
	\begin{align}
	\bcM({x'}+d{x'})&=\bM_{\rm slab}\,\bcM({x'})\nn\\
	&=\bcM({x'})-i\bcH({x'})\bcM({x'})d{x'}.
	\label{sch-diff}
	\end{align}
Because $x'$ is an arbitrary real number, the latter equation also holds if we change $x'$ to $x$. With this change of notation, {we can write \eqref{sch-diff} in the form, $id\bcM(x)=\bcH({x})\bcM({x})d{x}$, which is identical to the} ``time''-dependent Schr\"odinger equation,
	\be
	i\partial_x\bcM(x)=\bcH(x)\,\bcM(x).
	\label{sch-eq-x}
	\ee
Recalling that for ${x}\to-\infty$, $\hat\varepsilon({x})$ and $\hat\mu({x})$ tend to 1, we note that $\sS_{-\infty}$ represents the vacuum. Therefore, $\bcM(-\infty)=\bI$. This observation together with (\ref{sch-eq-U}) and (\ref{sch-eq-x}) show that $\bcM({x})$ coincides with the evolution operator $\bcU(x,x_0)$ for the matrix Hamiltonian (\ref{Ha=}) with initial ``time'' $x_0$ being set to $-\infty$, i.e., $\bcM(x)=\bcU(x,-\infty)$. Making use of this equation and the last relation in (\ref{limit}), we find
	\be
	\bM=\bcM(+\infty)=\bcU(+\infty,-\infty)=\sT\exp\left[-i\int_{-\infty}^\infty dx\:\bcH(x)\right].
	\label{M=exp2}
	\ee
	 
The above derivation of the Hamiltonian matrix $\bcH(x)$ whose evolution operator yields the transfer matrix for the TE and TM waves relies on the formula \eqref{M-slab} for the transfer matrix of a homogeneous slab and the (de)composition property (\ref{compose}). In the sequel, we offer an alternative derivation of $\bcH(x)$ which rests solely on the definition of the transfer matrix, namely \eqref{M=}.

Motivated by the approach pursued in Ref.~\cite{ap-2014} to obtain (\ref{H=1D}) and taking note of the fact that for every solution $\psi$ of \eqref{HH-eq-2}, $\psi$ and $\alpha^{-1}\partial_x\psi$ must be continuous functions of $x$, we introduce the two-component wave function,
	\be
	\bPsi:=\frac{1}{2}\left[\begin{array}{c}
	e^{-i\fK\,x}\{\psi-i(\fK\,\alpha)^{-1}\partial_x\psi\}\\[6pt]
	e^{i\fK\,x}\{\psi+i(\fK\,\alpha)^{-1}\partial_x\psi\}
	\end{array}\right].
	\label{Psi=}
	\ee
Because $\lim_{x\to\pm\infty}\alpha(x)=1$, {we can use (\ref{asymptot-psi-K}) and} (\ref{Psi=}) to show that
	\be
	\bPsi(\pm\infty)=\left[\begin{array}{c}
	A_\pm\\
	B_\pm\end{array}\right],
	\label{z1}
	\ee
where $\bPsi(\pm\infty):=\lim_{x\to\pm\infty}\bPsi(x)$. According to \eqref{M=} and (\ref{z1}), 
	\be
	\bPsi(+\infty)=\bM\bPsi(-\infty).
	\label{Psi-M-Psi}
	\ee 
This equation implies (\ref{M=exp2}), if we can find a matrix Hamiltonian $\bcH(s)$ such that 
	\be
	i\partial_x\bPsi(x)=\bcH(x)\bPsi(x).
	\label{z2}
	\ee
This assertion follows from the uniqueness of $\bM$ as the $2\times 2$ matrix fulfilling (\ref{Psi-M-Psi}) and not depending to $\bPsi(\pm\infty)$, and the fact that the evolution operator $\bcU(x,x_0)$ satisfies $\bPsi(+\infty)=\bcU(+\infty,-\infty)\bPsi(-\infty)$ and is independent of $\bPsi(\pm\infty)$.

Having obtained (\ref{z2}) we can determine the explicit form of $\bcH(x)$ by substituting (\ref{Psi=}) in this equation and using \eqref{HH-eq-2} to express its left-hand side in terms of $\psi$ and $\partial_x\psi$. Because the latter are linearly independent, we can solve the resulting equation for the entries of $\bcH(x)$. It is remarkable that this calculation reproduces the expression given by (\ref{Ha=}) for $\bcH(x)$.

We close this section by a simple application of Eqs.~(\ref{Ha=}) and (\ref{M=exp2}). 

\begin{figure}
        \begin{center}
        \includegraphics[scale=.2]{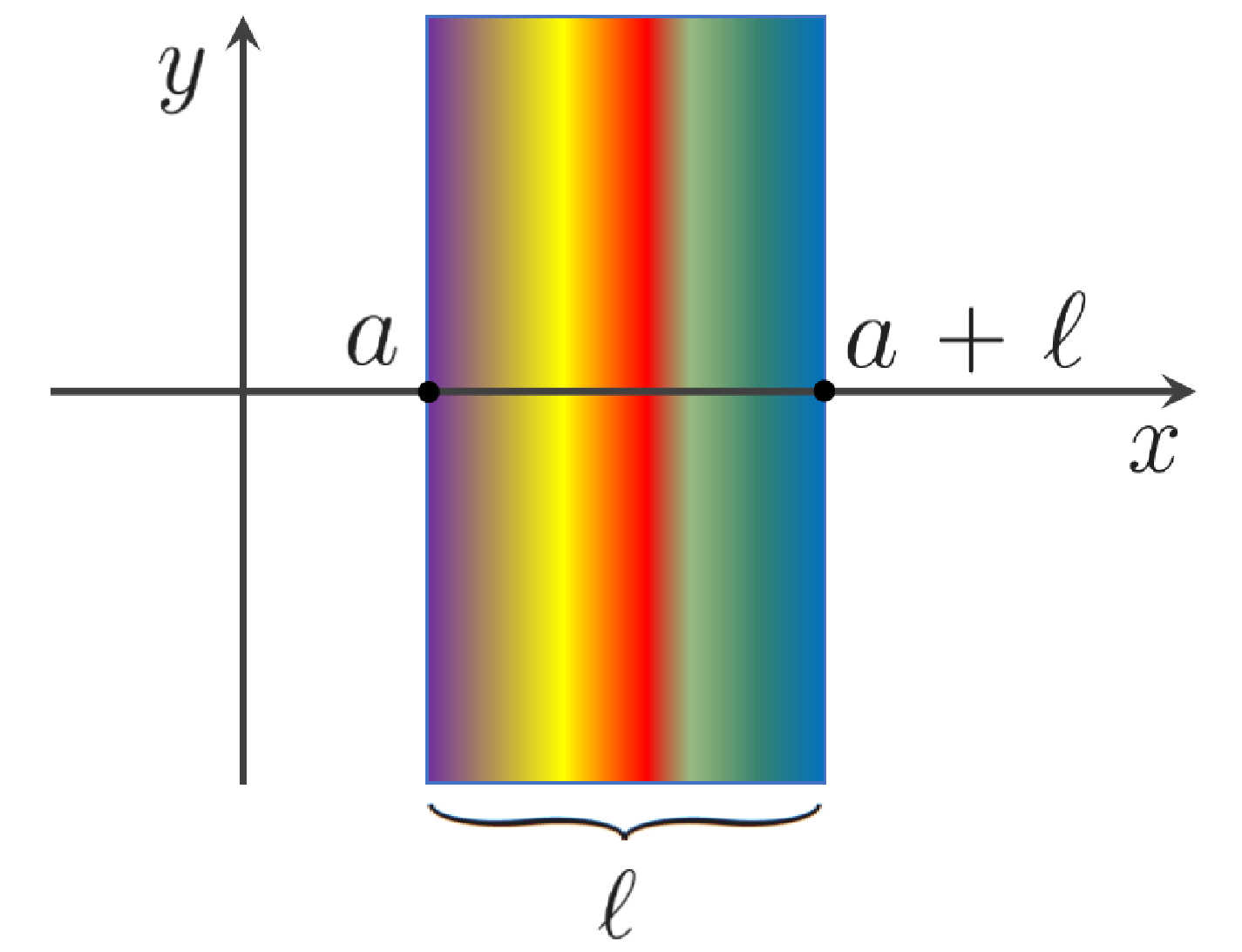}
        \caption{{Schematic views of an inhomogeneous slab with planar symmetry that lies between the planes $x=a$ and $x=a+\ell$.}}
        \label{fig4}
        \end{center}
        \end{figure}

Consider the cases where $\sS$ consists of a slab of thickness $\ell$ placed in vacuum, so that there is some $a\in\R$ such that $\hat\varepsilon(x)=\hat\mu(x)=1$ for $x\notin[a,a+\ell]$. {See Fig.~\ref{fig4}.}
Suppose that for $x\in[a,a+\ell]$, $\frac{\fn(x)^2-1}{\alpha(x)^2-1}$ takes a positive real constant value that is not greater than $1$. Then we can find an angle $\theta_\star\in (-90^\circ,90^\circ)$ fulfilling
	\be
	\cos\theta_\star=\sqrt{\frac{\fn(x)^2-1}{\alpha(x)^2-1}}.
	\label{condi}
	\ee
According to \eqref{Ha=} and \eqref{mmp=}, for $\theta=\theta_\star$ and $\theta=180^\circ-\theta_\star$,  $\fm_-(x)=0$, $\fm_+(x)=\alpha(x)$, $\bcH(x)$ is diagonal, and (\ref{M=exp2}) gives
	\be
	\bM=\exp\left[-i\int_{a}^{a+\ell}dx\:\bcH(x)\right]=
	\left[\begin{array}{cc}
	e^{ik\rho}& 0\\
	0 & e^{-ik\rho}\end{array}\right],
	\label{M-reflectionless}
	\ee
where 
	\[\rho:=\cos\theta_\star\int_{a}^{a+\ell}dx\:[\alpha(x)-1].\]
Comparing \eqref{M=S=} and (\ref{M-reflectionless}), we see that for incidence angles $\theta_\star$ and $180^\circ-\theta_\star$ {the reflection amplitudes vanish, i.e.,} the medium is reflectionless. If the slab is made of a nonmagnetic material, i.e., $\hat\mu=1$, we can satisfy (\ref{condi}) only for TM waves {and the incidence angle 
$\theta_\star=\arccos(\hat\varepsilon+1)^{-1/2}=
\arctan\fn$. This is the celebrated Brewster's angle.}
 
\section{Dynamical equations for reflection and transmission amplitudes}
\label{S4}

Consider the slab system $\sS$ of Fig.~\ref{fig4} where $\hat\varepsilon(x)=\hat\mu(x)=1$ for $x\notin[a,a+\ell]$, and let $\sS_{x'}$ be the corresponding truncated slab whose relative permittivity and permeability have the form \eqref{truncated-1}. Let $\cR^{l/r}(x')$ and $\cT(x')$ denote the left/right reflection and transmission amplitudes of $\sS_{x'}$, and $\bcM(x')$ be its transfer matrix. Then for all $x\in\R$,  
	\be
	\bcM(x)=\frac{1}{\cT(x)}\left[\begin{array}{cc}
	\cT(x)^2-\cR^l(x)\cR^r(x) & \cR^r(x)\\[6pt]
	-\cR^l(x) & 1\end{array}\right].
	\label{bcM}
	\ee
By construction, $\sS_x$ coincides with vacuum for $x\leq a$, and $\sS_{x}=\sS$ for $x\geq a+\ell$. In particular,
	\begin{align}
	&\bcM(a)=\bI
	&&\cR^{l/r}(a)=0,
	&&\cT(a)=1,
	\label{ini-condi}\\
	&\bcM(a+\ell)=\bM,
	&&\cR^{l/r}(a+\ell)=R^{l/r},
	&&\cT(a+\ell)=T.
	\label{full}
	\end{align}
	
{As we show in Appendix~A, substituting (\ref{Ha=}) and \eqref{bcM} in (\ref{sch-eq-x}), we find a set of three independent first order differential equations for $\cR^{l/r}$ and $\cT$ that are subject to the initial conditions (\ref{ini-condi}). More interestingly, we can} decouple these equations, reduce them to a single differential equation, namely
	\begin{align}
	&i\fK^{-1}\partial_x\cQ+\fm_-\cQ^2+2\fm_+\cQ+\fm_-=0,
	\label{Riccati}
	\end{align}
and show that
	\begin{align}
	&\cR^r(x)=e^{-2i\fK\, x}\cQ(x),
	\label{eq-R-Q}\\
	&\cT(x)=\exp\Big\{i\fK\int_a^{x}dx'\big[\fm_-(x')\cQ(x')+\fm_+(x')-1\big]\Big\},
	\label{eq-T-Q}\\
	&\cR^l(x)=i\fK\int_a^{x}dx'\:e^{2i\fK\,x'}\fm_-(x')\cT(x')^2.
	\label{eq-L-Q}
	\end{align}
These satisfy (\ref{ini-condi}) provided that $\cQ$ is the solution of the initial-value problem given by (\ref{Riccati}) and 
	\be
	\cQ(a)=0,
	\label{ini-Q}
	\ee
in the interval $[a,a+\ell]$.
	
In light of (\ref{full}), we can determine $R^{l/r}$ and $T$ by setting $x=a+\ell$ in (\ref{eq-R-Q}) -- (\ref{eq-L-Q}). This gives
	\begin{align}
	&R^r=e^{-2i\fK(a+\ell)}\cQ(a+\ell),
	\label{eq-R-Q-2}\\
	&T=\exp\Big\{i\fK\int_a^{a+\ell}dx\big[\fm_-(x)\cQ(x)+\fm_+(x)-1\big]\Big\},
	\label{eq-T-Q-2}\\
	&R^l=i\fK\int_a^{a+\ell}dx\:e^{2i\fK\,x}\,\fm_-(x)\cT(x)^2.
	\label{eq-L-Q-2}
	\end{align}
Recalling that according to \eqref{mmp=}, $\fm_-(x)=0$ for $x\notin[a,a+\ell]$, we can express \eqref{eq-L-Q-2} as
	\begin{align}
	&R^l=i\fK\int_{-\infty}^{\infty}dx\:e^{2i\fK\,x}\fm_-(x)\cT(x)^2=i\fK \tilde F(-2\fK),
	\nn
	\end{align}
where $\tilde F(p)$ denotes the Fourier transform of the function, $F(x):=\fm_-(x)\cT(x)^2$,
i.e., $\tilde F(p):=\int_{-\infty}^\infty dx\:e^{-ip x}F(x)$.

{Equations (\ref{eq-R-Q-2}) -- (\ref{eq-L-Q-2}) provide a novel method of solving the scattering problem for TE and TM waves which reduces it to the initial-value problem for a first order differential equation, namely the one given by (\ref{Riccati}) and (\ref{ini-Q}). This method involves the following steps.
	\begin{enumerate}
	\item Obtain the solution $\cQ(x)$ of (\ref{Riccati}) that satisfies the initial condition (\ref{ini-Q}).
	\item Substitute $\cQ(x)$ in (\ref{eq-T-Q}) to determine $\cT(x)$.
	\item Insert $\cQ(x)$ and $\cT(x)$ in (\ref{eq-R-Q}) -- (\ref{eq-L-Q}) to find $R^{l/r}$ and $T$.
	\end{enumerate}
We wish to emphasize that the reduction of the scattering problem to an initial-value problems for a first-order differential equation is of practical importance, because the latter admits a straightforward numerical solution whenever a solution exists.}{\footnote{{It is actually easy to carry out all three steps of the above method using Mathematica or Maple.}}} {Because (\ref{Riccati}) is a nonlinear equation, it may admit blow-up solutions, i.e., there may exist $\ell$ for which $\cR(a+\ell)=\infty$. This happens precisely
at a spectral singularity \cite{prl-2009} which corresponds to the onset of lasing \cite{pra-2011a,longhi-2010,prsa-2012,jo-2017}.}

{Because (\ref{Riccati}) is a Riccati equation, for generic choices of $\fm_-$, we can reduce it to a second-order linear homogeneous equation. This shows that we can reduce the scattering problem to the initial-value problem for such a second-order linear differential equation. We present the details of this procedure in Appendix~B.}

\section{Unidirectional reflectionlessness for TE and TM waves}
\label{S5}

According (\ref{eq-R-Q-2}) the right reflection amplitude of the slab $\sS$ vanishes if and only if $\cQ(a+\ell)=0$. This observation suggests that we can identify permittivity and permeability profiles displaying reflectionlessness for right-incident TE and TM waves by choosing a differentiable function $\cQ:[a,a+\ell]\to\C$ that satisfies
	\be
	\cQ(a)=\cQ(a+\ell)=0,
	\label{condi-3}
	\ee
substituting it in (\ref{Riccati}), and using the resulting equation together with (\ref{mmp=}) to determine $\hat\varepsilon$ or $\hat\mu$.

{To arrive at a quantitative description of this method, we first note that according to (\ref{mmp=}),
	\be
	\fm_+(x)=\fm_-(x)+\alpha(x).
	\label{mm=mp-a}
	\ee
This allows us to express (\ref{Riccati}) in the form}
	\be
	i\fK^{-1}\partial_x\cQ+\fm_-(\cQ+1)^2+2\alpha\,\cQ=0.
	\label{Riccati-2}
	\ee
Next, we introduce
	\be
	\beta:=\left\{\begin{array}{cc}
	\hat\varepsilon &\mbox{for TE waves},\\
	\hat\mu&\mbox{for TM waves},\end{array}\right.
	\label{beta}
	\ee	
and use (\ref{mmp=}) to show that
	\be
	\fm_-=\frac{1}{2}\left[\sec^2\theta(\beta-\alpha^{-1})+\alpha^{-1}-\alpha\right].
	\label{fmm=2}
	\ee
Substituting this equation in (\ref{Riccati-2}), solving for $\beta$, and noting that $\fK:=k|\cos\theta|$, we find
	\be
	\beta(x)=\frac{\sin^2\theta}{\alpha(x)}+\cos^2\theta\left\{\left[\frac{\cQ(x)-1}{\cQ(x)+1}\right]^2\alpha(x)-\frac{2i\partial_x\cQ(x)}{k|\cos\theta|\, [\cQ(x)+1]^2}\right\},
	\label{s5-e1}
	\ee
where $x\in[a,a+\ell]$. We can alternatively view (\ref{s5-e1}) as an equation for $\alpha(x)$. Solving this equation we find,
	\be
	\alpha(x)=\xi\pm\sqrt{\xi(x)^2-\zeta(x)^2},
	\label{s5-e2}
	\ee
where again $x\in[a,a+\ell]$, and 
	\begin{align}
	&\xi:=\frac{2i|\cos\theta|\,\partial_x\cQ+k\beta(\cQ+1)^2}{2k\cos^2\theta\,(\cQ-1)^2},
	&&\zeta:=\frac{\tan\theta(\cQ+1)}{\cQ-1}.
	\label{xi-zeta}
	\end{align}
Note that in view of (\ref{alpha=}) and (\ref{beta}), $\alpha(x)=\beta(x)=1$ for $x\notin[a,a+\ell]$. This shows that (\ref{s5-e1}) determines $\hat\varepsilon$ in terms of $\hat\mu$ and $\cQ$ for TE waves and $\hat\mu$ in terms of $\hat\varepsilon$ and $\cQ$ for TM waves. Similarly, (\ref{s5-e2}) specifies $\hat\mu$ in terms of $\hat\varepsilon$ and $\cQ$ for TE waves and $\hat\varepsilon$ in terms of $\hat\mu$ and $\cQ$ for TM waves. 

{We wish to stress that the $k$ and $\theta$ appearing in (\ref{s5-e1}) and (\ref{xi-zeta}) are respectively the wavenumber and incidence angle for which our slab displays right-reflectionlessness for TE or TM waves. In the following we use $k_\star$ and $\theta_\star$ for this wavenumber and incidence angle to distinguish them from generic wavenumbers and incidence angles. In other words, the right-reflectionlessness occurs for $k=k_\star$ and $\theta=\theta_\star$.\footnote{{Recall that for a right-incidant wave, $90^\circ<\theta_\star<270^\circ$.}}}
	
As a special case, consider TE waves scattered by a nonmagnetic slab, so that $\alpha=\hat\mu=1$ and $\beta=\hat\varepsilon$. {Then (\ref{mm=mp-a}) and (\ref{fmm=2}) give
	\be
	\fm_-=\fm_+-1=\frac{\hat\varepsilon-1}{2\cos^2\theta},
	\label{m-minus=}
	\ee
and (\ref{s5-e1}) with $k=k_\star$ and $\theta=\theta_\star$ becomes
	\be
	\hat\varepsilon(x)=1-2\cos^2\theta_\star\left\{\frac{i\fK_\star^{-1}\partial_x\cQ(x)+2\cQ(x)}{[\cQ(x)+1]^2}\right\}\chi_{a,a+\ell}(x),
	\label{eps=TE}
	\ee
where 
	\[\fK_\star:=k_\star|\cos\theta_\star|,\]}%
and for all $a,b\in\R$ with $a<b$,
	\[\chi_{a,b}(x):=\left\{\begin{array}{cc}
	1 &\for~a\leq x\leq b,\\
	0 &{\rm otherwise}.\end{array}\right.\]

{For a slab possessing a permittivity profile of the form (\ref{eps=TE}), $R^r=0$ for $k=k_\star$. We can also use (\ref{eq-T-Q}), (\ref{eq-T-Q-2}), (\ref{eq-L-Q-2}), (\ref{m-minus=}), and (\ref{eps=TE}) to derive the following expressions for its transmission and left reflection amplitudes at $k_\star$.
	\begin{align}
	T&=e^{2i\fK_\star[\Delta(\ell)-\ell]},
	\label{T=TE}\\
	R^l&=\int_a^{a+\ell} dx\:[\partial_x\cQ(x)-2i\fK_\star\cQ(x)]\,
	e^{2i\fK_\star[2\Delta(x)-x+2a]}\nn\\
	&=-4i\fK_\star\int_{a}^{a+\ell}dx\left\{\frac{\cQ(x)\,e^{2i\fK_\star[2\Delta(x)-x+2a]}}{\cQ(x)+1}
	\right\},
	\label{L=TE}
	\end{align}
where
	\be
	\Delta(x):=\int_a^{x}\frac{dx'}{\cQ(x')+1},
	\label{Delta=}
	\ee
and we have also made use of (\ref{condi-3}).}

{If $\cQ$ happens to be a real-valued function, $\Delta$ takes real values as well, and $T$ is a phase factor. This is a rather nontrivial result. We can justify it when $\cQ$ satisfies 
	\be
	\cQ(\ell-x)^*=\cQ(x).
	\label{PT}
	\ee
This condition marks the $\cP\cT$-symmetry of the slab with $\cP$ and $\cT$ respectively denoting the space reflection about the plane $x=\ell/2$, i.e., $x\to\ell-x$, and complex-conjugation. If (\ref{PT}) holds, the permittivity profile (\ref{eps=TE}) is also $\cP\cT$-symmetric, and its reflection and transmission amplitudes fulfill the generalized unitarity condition,  $|T|^2\pm|R^lR^r|=1$, \cite{Springer-book-2018,Ge}. Because $R^r=0$, this condition implies $|T|=1$. In general, $\cQ$ can be a real and non-$\cP\cT$-symmetric function. In this case $\hat\varepsilon$ need not be $\cP\cT$-symmetric, yet the above analysis shows that $|T|=1$.}
	
{If we take an arbitrary differentiable function $\cQ:[a,a+\ell]\to\C$ fulfilling (\ref{condi-3}) and substitute it} in (\ref{eps=TE}), we find the relative permittivity of a slab that does not reflect right-incident TE waves with wavenumber $k=k_\star$ and incidence angle $\theta=\theta_\star$. As an example, consider taking 
	\begin{align}
	&a=0,
	&&\cQ(x)=\fz\,\sin({K}_nx),
	&&{K}_n:=\frac{\pi n}{\ell},
	\nn
	\end{align}
where $\fz$ is a possibly complex constant, and $n$ is an integer. Then (\ref{eps=TE}) gives
	{\be
	\hat\varepsilon(x)=1-4\fz\cos^2\theta_\star\left\{\frac{\sin({K}_nx)+i({K}_n/2\fK_\star)\cos({K}_nx)}{[1+\fz\sin({K}_nx)]^2}\right\}\chi_{0,\ell}(x).
	\label{eps=TE-eq1}
	\ee}
		
{For $\fK_\star=K_n/2$, i.e.,
	\be
	k_\star=\frac{{K}_n}{2|\cos\theta_\star|}=\frac{\pi n}{2\ell|\cos\theta_\star|},\nn
	\ee
Eq.~(\ref{eps=TE-eq1}) becomes
	\be
	\hat\varepsilon(x)=1-\frac{4i\fz\cos^2\theta_\star\,e^{-i{K}_n x}\:\chi_{0,\ell}(x)}{[1+\fz\sin({K}_nx)]^2}.
	\label{eps=TE-eq1-2}
	\ee
Because for a nonmagnetic material, $\fn^2=\hat\varepsilon$, the potential (\ref{optical-v}) takes the form,}
	\be
	v(x)=\frac{i\fz {K}_n^2\,e^{-i{K}_n x}\;\chi_{0,\ell}(x)}{[1+\fz\sin({K}_nx)]^2}.
	\label{exp-pot}
	\ee
For $|\fz|\ll 1$, we can ignore the term $\fz\sin({K}_nx)$ on the right-hand sides of (\ref{eps=TE-eq1-2}) and (\ref{exp-pot}). This yields,
 	\begin{align}
	&\hat\varepsilon(x)\approx1-4i\fz\cos^2\theta_\star\, e^{-i{K}_n x}\chi_{0,\ell}(x),
	&&v(x)\approx i\fz {K}_n^2\,e^{-i{K}_n x}\chi_{0,\ell}(x).
	\label{approx}
	\end{align}
The slab systems described by these relations generalize the model considered in Ref.~\cite{lin} in the context of unidirectional invisibility to the scattering of oblique TE waves.{\footnote{{See also Refs.~\cite{poladian,greenberg,kulishov,jpa-2016}.}}} Notice that unlike the systems specified by (\ref{approx}) which display approximate reflectionlessness \cite{longhi-2011}, the permittivity profile (\ref{eps=TE-eq1-2}) is exactly reflectionless for the specified right-incident TE waves. 
	
Another simple choice for $\cQ(x)$ is
	\be
	\cQ(x)=\kappa^2 x(\ell-x),
	\label{Q=eg2}
	\ee 
where $\kappa$ is a positive real constant having the dimension of length$^{-1}$. {Setting $a=0$ and substituting (\ref{Q=eg2}) in (\ref{eps=TE}), we obtain
	\be
	\hat\varepsilon(x)=1-2\kappa^2\cos^2\theta_\star\left\{
	\frac{2x(\ell-x)+i \fK_\star^{-1}(\ell-2x)}{
	 [\kappa^2 x(\ell-x)+1]^2}\right\}\chi_{0,\ell}(x).
	\label{eps=TE-eg2}
	\ee
This is another example of a permittivity profile that does not reflect a TE wave with wavenumber $k_\star$ and incidence angle $\theta_\star\in(90^\circ,270^\circ)$. Notice that unlike (\ref{eps=TE-eq1-2}), it fails to be locally periodic. Furthermore, because we take $\kappa$ to be a real parameter, $\cQ$ is real-valued. It also satisfies $\cQ(\ell-x)=\cQ(x)$. Therefore (\ref{PT}) holds, and the slab is $\cP\cT$-symmetric.}

{For the choice of $\cQ$ given by (\ref{Q=eg2}), we can evaluate the integral on the right-hand side of (\ref{Delta=}) analytically. The results is 
	\be
	\Delta(x)=\frac{1}{\kappa \sqrt{(\kappa\ell)^2+4}}\ln\left\{\frac{
	\kappa x\left[\kappa\ell+\sqrt{(\kappa\ell)^2+4}\right]+2}{
	\kappa x\left[\kappa\ell-\sqrt{(\kappa\ell)^2+4}\right]+2}\right\}.
	\label{Delta=eg2}
	\ee
Substituting this equation in (\ref{T=TE}) and (\ref{L=TE}), we find the transmission amplitude and the left reflection amplitude of the permittivity profile (\ref{eps=TE-eg2}) for a TE wave with wavenumber $k_\star$ and incidence angle $\theta_\star$. Because $\cQ$ is real-valued, the transmission amplitude is a phase factor. We can express it in the form, $T=e^{-2i\fK_\star\ell\,\varphi}$, where $\varphi:=1-\Delta(\ell)/\ell$. This means that our slab serves as a phase shifter for TE waves with wavenumber $k_\star$ and incidence angle $\theta_\star$.\footnote{{In view of transmission reciprocity \cite{tjp-2020,Springer-book-2018}, the same holds for left-incident TE waves with wavenumber $k_\star$ and incidence angle $180^\circ-\theta_\star$.}}}
 
{As seen from (\ref{Delta=eg2}), $\varphi$ is a real-valued function of the dimensionless parameter $\kappa\ell$. In view of (\ref{eps=TE-eg2}), for fixed $\ell$, this parameter is a measure of the strength of the scattering effects of our slab. Figure~\ref{fig5} provides a graphical demonstration of the dependence of $\varphi$ on $\kappa\ell$. 
	\begin{figure}
        \begin{center}
        \includegraphics[scale=.21]{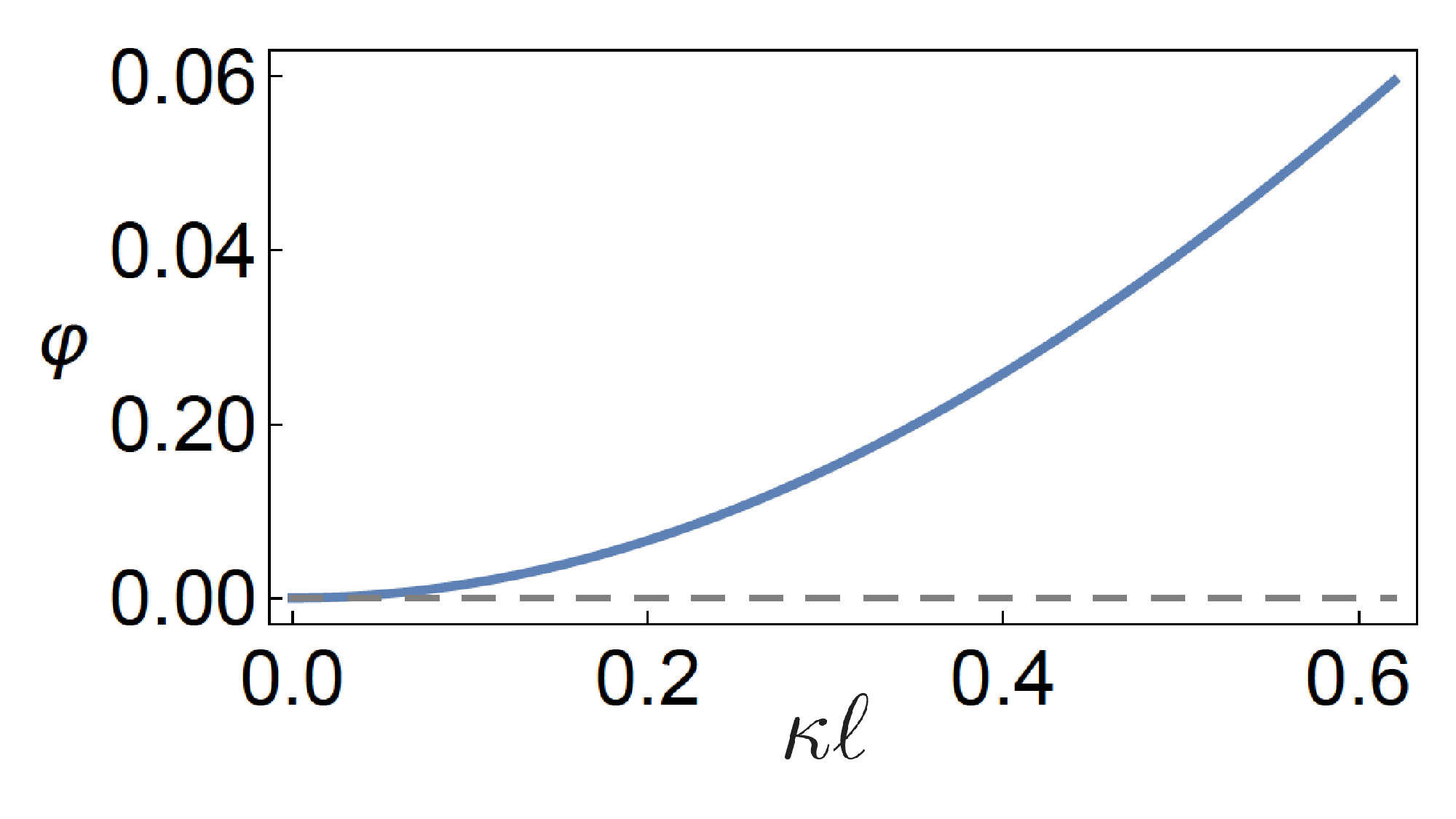}~~~~~~~~~~~~~
        \includegraphics[scale=.21]{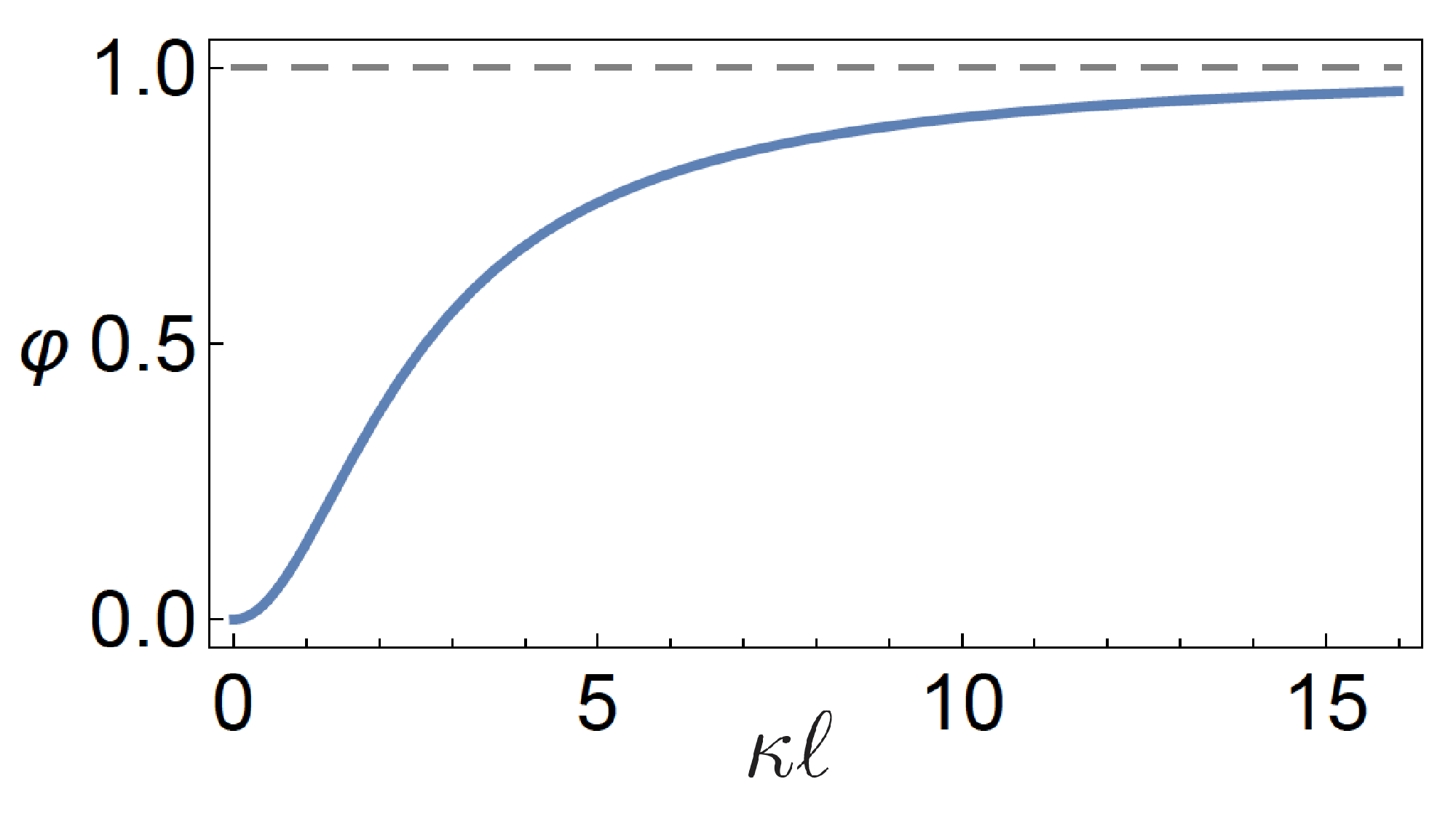} 
        \caption{{Plots of $\varphi$ as a function of $\kappa\ell$ for different ranges of values of the latter. For $\kappa\ell\to 0$ and $\kappa\ell\to\infty$, $\varphi$ tends to $0$ and $1$, respectively. The dashed lines represent the limiting values of $\varphi$.}}
        \label{fig5}
        \end{center}
        \end{figure}%
For $\kappa\ell\gg 1$, $\phi\approx1$, and $T\approx e^{-2i\fK_\star\ell}$. For $\kappa\ell\ll 1$,  $\varphi=\frac{1}{6}(\kappa\ell)^2+\cO(\kappa\ell)^3$, and consequently
	\begin{align}
	&T=1-\frac{i}{3}\fK_\star\ell(\kappa\ell)^2+\cO(\kappa\ell)^4,
	\label{weak}
	\end{align}
where $\cO(\kappa\ell)^n$ stands for the terms of order $n$ and higher in powers of $\kappa\ell$.
According to \eqref{weak}, if $\fK_\star\ell(\kappa\ell)^2$ is negligibly small, our slab is approximately transparent and therefore invisible from the right. To decide whether this approximate invisibility is unidirectional, we need to explore $R^l$.}

{Because the integral on the right-hand side of (\ref{L=TE}) cannot be evaluated, we do not have an analytic expression for $R^l$. We can, however, use (\ref{L=TE}), (\ref{Q=eg2}), and the fact that $\cQ$ takes real and nonnegative values to establish,
	\be
	|R^l|\leq 4\fK_\star\kappa^2\int_0^\ell dx\left[\frac{x(x-\ell)}{\kappa^2x(x-\ell)+1}\right]
	\leq 4\fK_\star\kappa^2\int_0^\ell dx\, x(x-\ell)=\frac{2\fK_\star\ell(\kappa\ell)^2}{3}.
	\nn
	\ee
This relation shows that, for $\kappa\ell\ll 1$ and $\fK_\star\ell(\kappa\ell)^2\ll 1$, the permittivity profile (\ref{eps=TE-eg2}) is approximately reflectionless from the left. Therefore its approximate invisibility is bidirectional. Note however that if $\kappa\ell\ll 1$ but $\fK_\star\ell$ is so large that $\fK_\star\ell(\kappa\ell)^2$ is no longer negligible, $T\neq 1$. This means that the slab is not transparent and hence cannot be invisible from either left or right.}

{Figure~\ref{fig6} shows plots of $|R^l|$ for different values of $\kappa\ell$ and $\fK_\star\ell$. As seen from these plots,  for sufficiently large values of $\fK_\star\ell$, $|R^l|$ is not negligible. Because $R^r=0$ for arbitrary values of $\kappa\ell$ and $k_\star\ell$, the slab displays exact unidirectional reflectionlessness.
\begin{figure}
        \begin{center}
        \includegraphics[scale=.35]{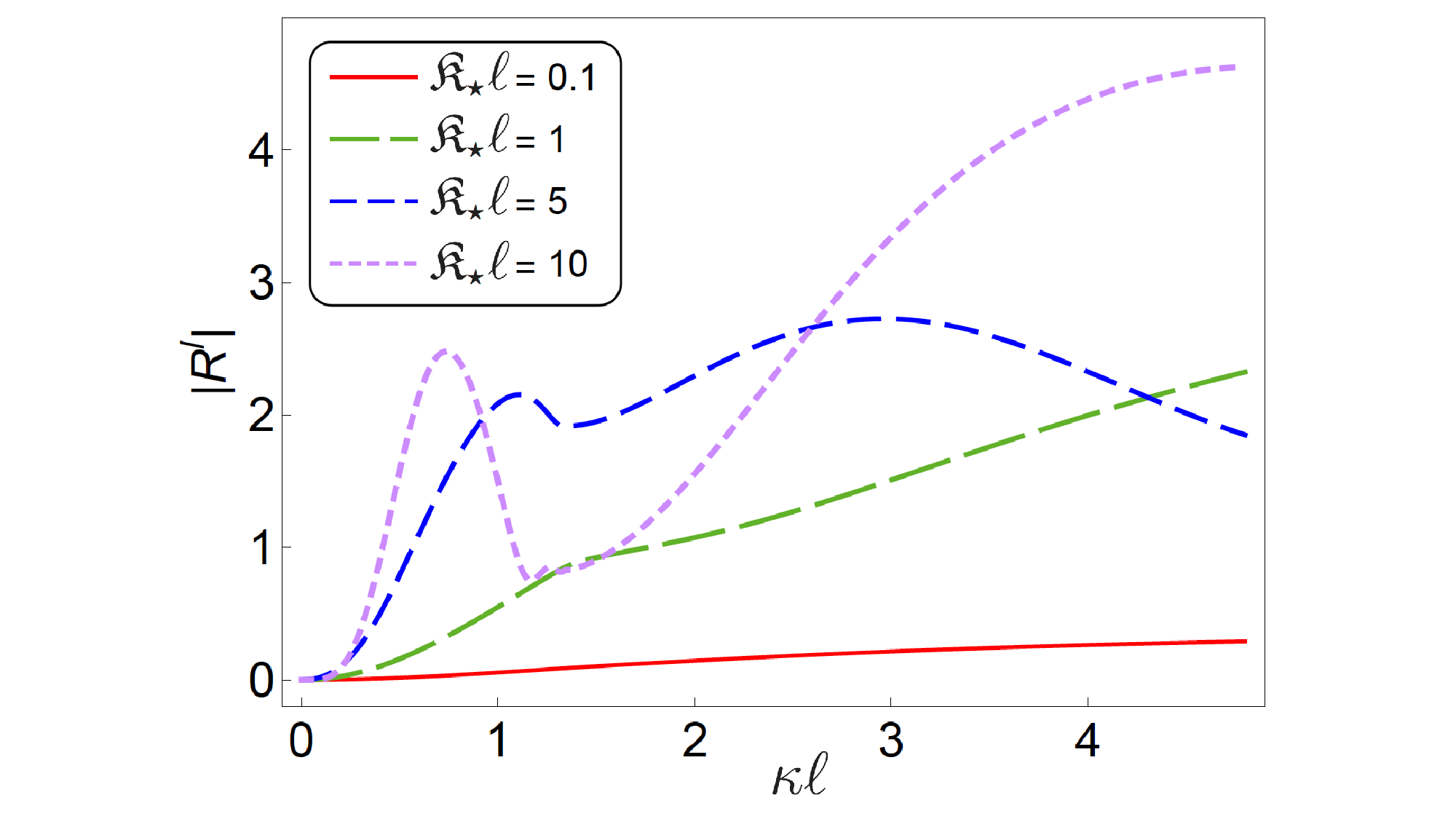}
        \caption{{Plots of $|R^l|$ as a function of $\kappa\ell$ for $k=k_\star$, $\theta=\theta_\star$ and different values of $\fK_\star\ell$.}}
        \label{fig6}
        \end{center}
        \end{figure}%
}

Changing $\hat\varepsilon$ to $\hat\mu$, and then setting $\hat\varepsilon=1$ in the above analysis, we can also find purely magnetic permeability profiles that display reflectionlessness from right for TM waves with a specific wavenumber and incidence angle. 

We can also use the above method to obtain slabs that are reflectionless from the left. This is simply because a system determined by $\hat\varepsilon$ and $\hat\mu$ is left-reflectionless, i.e., $R^l=0$ for an incident wave with wavenumber $k_\star$ and incidence angle $\theta_\star$ if and only if the time-reversed system given by the relative permittivity $\hat\varepsilon^*$ and relative permeability $\hat\mu^*$ is reflectionless from the right {for the same wavenumber and the incidence angle $180^\circ-\theta_\star$.} This follows from the transformation property of the transfer matrix under time-reversal transformation \cite{Springer-book-2018}.

\section{Concluding remarks}
\label{S6}

Transfer matrix is a powerful tool for conducting scattering calculations particularly when dealing with multi-layer and locally periodic scatterers. In practice its use involves dissecting the scatterer into sufficiently thin slices whose transfer matrices are easier to calculate. The transfer matrix of the scatterer is then obtained by multiplying the transfer matrices of the slices in a particular order. Proceeding in the opposite direction, we can actually derive a general Dyson series expansion of the transfer matrix of the scatterer by identifying it with $\bcU(-\infty,\infty)$, where $\bcU(x,x_0)$ is the time-evolution operator for a fictitious non-unitary quantum system. This leads to a dynamical formulation of potential scattering with a variety of interesting applications \cite{ap-2014,pra-2021}. 

In the present article we developed a similar approach for dealing with the scattering of TE and TM waves by the inhomogeneities of a general isotropic linear medium possessing planar symmetry. The transfer matrix describing the scattering of these waves turns out to admit a Dyson series expansion defined by a non-Hermitian Hamiltonian. This shows that the scattering phenomenon for these waves is intimately linked with the quantum dynamics generated by non-Hermitian Hamiltonians.

A direct implication of this observation is the existence of dynamical equations for the reflection and transmission amplitudes of TE and TM waves. We could decouple these equations and reduce them to a single Riccati equation (alternatively a second-order linear homogeneous ordinary differential equation.) For cases where the inhomogeneity of the medium is confined to an infinite planar slab, we reduced the solution of the scattering problem to that of an initial-value problem for this equation which can easily be obtained numerically for example using Mathematica or Maple. 

Another interesting application of the dynamical equation determining the reflection and transmission amplitudes is that it provides a very simple scheme for identifying slabs that do not reflect TE or TM waves with a given wavenumber and incidence angle. This provides an effective method of generating permittivity and permeability profiles that display exact (nonperturbative) reflectionlessness for any TE or TM wave.

{To the best of our knowledge the approach outlined in this article is the first to offer a method for mapping the scattering problem for TE and TM waves to an initial-value problem for a first-order differential equation. It is also the first to offer a systematic prescription for generating unidirectionally reflectionless permittivity profiles for TE and TM waves with arbitrary incidence angles and wavenumbers.} 

{Finally, we wish to note that because Eq.~\eqref{HH-eq-2} models the propagation of pressure waves in a compressible fluid with planar symmetry \cite{Bergmann,Martin}, our results provide a dynamical formulation of acoustic scattering in one dimension. In particular, they can be used to turn these scattering problems into easily solvable first-order initial-value problems and identify configurations of such fluids that display nonreciprocal reflection for certain sound waves.}

\subsection*{Acknowledgements}
This work has been supported by the Scientific and Technological Research Council of T\"urkiye (T\"UB\.{I}TAK) in the framework of the project 123F180 and by Turkish Academy of Sciences (T\"UBA). {We are indebted to Prof.\ Vladimir Konotop for bringing Ref.~\cite{Klyatskin} to our attention.}

\subsection*{Appendix~A: Differential equations for $\cR^{l/r}$ and $\cT$ and their solution}

Substituting (\ref{Ha=}) in (\ref{sch-eq-x}) yields the following system of equations for the entries of $\bcM(x)$.
	\begin{align}
	&i\fK^{-1}\partial_x\cM_{11}=(1-\fm_+)\cM_{11}-e^{-2i\fK\, x}\fm_-\cM_{21},
	\label{app-11}\\
	&i\fK^{-1}\partial_x\cM_{12}=(1-\fm_+)\cM_{12}-e^{-2i\fK\, x}\fm_-\cM_{22},
	\label{app-12}\\
	&i\fK^{-1}\partial_x\cM_{21}=e^{2i\fK\, x}\fm_-\cM_{11}-(1-\fm_+)\cM_{21},
	\label{app-21}\\
	&i\fK^{-1}\partial_x\cM_{22}=e^{2i\fK\, x}\fm_-\cM_{12}-(1-\fm_+)\cM_{22}.
	\label{app-22}
	\end{align}
If we use (\ref{bcM}) to express $\cM_{ab}$ in terms of $\cR^{l/r}$ and $\cT$ and plug the result in (\ref{app-12}) -- (\ref{app-22}), we respectively find
	\begin{align}
	&i\fK^{-1}\cT\partial_x\left(\frac{\cR^r}{\cT}\right)=(1-\fm_+)\cR^r-e^{-2i\fK\, x}\fm_-,
	\label{app-12-z}\\
	&i\fK^{-1}\cT\partial_x\left(\frac{\cR^l}{\cT}\right)=-e^{2i\fK\, x}\fm_-(\cT^2-\cR^l\cR^r)+
	(\fm_+-1)\cR^l,
	\label{app-21-z}\\
	&i\fK^{-1}\cT\partial_x\left(\frac{1}{\cT}\right)=e^{2i\fK\, x}\fm_-\cR^r+\fm_+-1.
	\label{app-22-z}
	\end{align}
Integrating both sides of (\ref{app-22-z}) and imposing the condition, $\cT(a)=1$, we obtain
	\be
	\cT(x)=\exp\Big\{i\fK\int_a^{x}dx'[e^{2i\fK\, x'}\fm_-(x')\cR^r(x')+\fm_+(x')-1]\Big\}.
	\label{eq-T}
	\ee

Next, we expand $\partial_x(\cR^{l/r}/\cT)$ in terms of $\partial_x\cR^{l/r}$ and $\partial_x\cT$, use (\ref{app-22-z}) to express the latter in terms of $\cR^r$, and substitute the result in (\ref{app-12-z}) and (\ref{app-21-z}). This gives
	\begin{align}
	&i\fK^{-1}\partial_x\cR^r+e^{2i\fK\, x}\fm_-{\cR^{r}}^2+2(\fm_+-1)\cR^r+e^{-2i\fK\, x}\fm_-=0,
	\label{eq-R}\\
	&i\fK^{-1}\partial_x\cR^l=-e^{2i\fK\, x}\fm_-\cT^2.
	\label{eq-L-z}
	\end{align}
Integrating both sides of the latter equation we find (\ref{eq-L-Q}). Introducing $\cQ(x):=e^{2i\fK\, x}\cR^r(x)$, which satisfies (\ref{eq-R-Q}), and substituting this equation in (\ref{eq-R}) and (\ref{eq-T}) we are led to (\ref{eq-R-Q}) and (\ref{eq-T-Q}), respectively.

{For the special case of a normally incident TE wave scattered by a nonmagnetic slab, where $\alpha=\cos\theta=1$, $\fK=k$, and $\fm_-=\fm_+-1=\frac{1}{2}(1-\hat\varepsilon)$, Eq.~(\ref{eq-R}) becomes
	\be
	\partial_x\cR^r=\frac{ik}{2}(\hat\varepsilon-1)\left(
	e^{ik x}{\cR^{r}}+e^{-ik x}\right)^2.
	\nn
	\ee
This coincides with Eq.~(18) of Ref.~\cite{ap-2014} which is derived for the right reflection amplitude of the potential, $v=k^2(1-\hat\varepsilon)$, as well as Eq.~(1.18) of Ref.~\cite{Klyatskin} where the author presents it as an application of the so-called embedding method of stochastic analysis.\footnote{The symbol $L$, $\varepsilon$, and $R_L$ of Ref.~\cite{Klyatskin} respectively correspond to $x$, $\hat\varepsilon-1$, and $\cQ(x)$ of the present article.}}

\subsection*{{Appendix~B: Reduction of the scattering problem to a second-order linear equation}}

{Equation (\ref{Riccati}) is a Riccati equation. This suggests that we can reduce it to a second-order linear homogeneous equation provided that, for all $x\in[a,a+\ell]$, $\fm_-(x)\neq 0$ and $\partial_x\fm_-(x)$ exists.  
For example, introducing
	\be
	\cX(x):=\exp\Big[-i\fK\int_a^{x}dx'\:\fm_-(x')\cQ(x')\Big],
	\label{X=}
	\ee
we can check that (\ref{Riccati}) and (\ref{ini-Q}) are equivalent to
	\be
	\partial_x^2\cX-(\partial_x\ln\fm_-+2i\fK\,\fm_+)\partial_x\cX-\fK^2\fm_-^2\cX
	=0.
	\label{eq-X}
	\ee
	}%
	
{Notice that according to (\ref{ini-Q}) and (\ref{X=}), $\cX$ fullfils the initial conditions,
	\begin{align}
	&\cX(a)=1,
	&&\partial_x\cX(a)=0.
	\label{S-ini-condi}
	\end{align}
Furthermore, we can use (\ref{eq-R-Q}) -- (\ref{eq-L-Q}) and (\ref{X=}) to show that
	\begin{align}
	&\cR^r(x)=\frac{ie^{-2i\fK\,x}\partial_x \cX(x)}{\fK\,\fm_-(x)\, \cX(x)},
	\quad\quad\quad 
	\cT(x)=\frac{\eta(x)}{\cX(x)},
	\label{RT=R}\\[6pt]
	&\cR^l(x)=i\fK \int_a^{x}dx'\: 
	\frac{e^{2i\fK\, x'}\fm_-(x')\,\eta(x')^2}{\cX(x')^2},
	\label{R=L}
	\end{align}
where
	\[\eta(x):=\exp\Big\{i\fK\int_{a}^xdx'\:[\fm_+(x')-1]\Big\}.\]	
Substituting (\ref{eq-L-Q}), (\ref{RT=R}), and (\ref{R=L}) in (\ref{full}), we have
	\begin{align}
	&R^r=\frac{ie^{-2i\fK(a+\ell)}\partial_x \cX(a+\ell)}{\fK\,\fm_-(a+\ell)\, \cX(a+\ell)},
	\quad\quad\quad T=\frac{\eta(a+\ell)}{\cX(a+\ell)},
	\label{Sol-RT}\\[6pt]
	&R^l=i\fK \int_a^{a+\ell}dx\:
	\frac{e^{2i\fK\, x}\fm_-(x)\eta(x)^2}{\cX(x)^2}.
	\label{Sol-L}
	\end{align}  
	}%
		
{Equations (\ref{Sol-RT}) and (\ref{Sol-L}) reduce the solution of the scattering problem for the slab we consider in Sec.~\ref{S4} to an initial-value (dynamical) problem for a second-order homogeneous differential equation in the interval $[a,a+\ell]$, namely the one given by (\ref{eq-X}) and (\ref{S-ini-condi}). In light of (\ref{mm=mp-a}), the solution of this problem is uniquely determined by the choice of $\alpha(x)$ and $\fm_-(x)$ with the latter  assumed to be nonzero and differentiable inside the slab. If there are $x_1,x_2,\cdots,x_n\in(a,a+\ell)$ where this condition fails, we can dissect $[a,a+\ell]$ into subintervals where we can employ this method to determine the reflection and transmission amplitudes, use (\ref{M=S=}) to find the transfer matrix for each of the corresponding slices of the slab, and then employ the composition property (\ref{compose}) to obtain its transfer matrix. Using the result of this calculation together with (\ref{R=}) and (\ref{T=})  we can then determine the reflection and transmission amplitudes of the slab. Alternatively, we can employ the method outlined in Sec.~\ref{S4} which is based on the solution of the initial-value problem given (\ref{Riccati}) and (\ref{ini-Q}).}

\ed